\documentclass[3p,times,review,numbers,authoryear]{elsarticle}

\usepackage{graphicx}
\usepackage{caption}
\usepackage{subfig}
\usepackage{array}
\usepackage{mathtools}
\usepackage[mathscr]{euscript}
\usepackage{enumitem}
\usepackage{float}
\usepackage{epstopdf}
\usepackage{algorithm}
\usepackage{stmaryrd}
\usepackage{algpseudocode}
\usepackage{multirow}
\usepackage{booktabs}
\usepackage{url}
\usepackage{hhline}
\usepackage{color} 
\usepackage{amssymb}
\usepackage[numbers]{natbib}

\newtheorem{thm}{Theorem}

\newdefinition{rmk}{Remark}
\newproof{pf}{Proof}
\begin{document}

\begin{frontmatter}
This is a preprint of the paper:

David Meg\'{i}as, Amna Qureshi, ``Collusion-resistant and privacy-preserving P2P multimedia distribution based on recombined fingerprinting", Expert Systems with Applications, Volume 71, $1^{\mathit{st}}$ April 2017, Pages 147-172, ISSN 0957-4174, \url{http://dx.doi.org/ 10.1016/j.eswa.2016.11.015}.

\newpage

\title{Collusion-resistant and privacy-preserving P2P multimedia distribution based on recombined fingerprinting}


\author{David Meg\'{i}as, Amna Qureshi}
\address{Estudis d'Inform\`{a}tica Multim\`{e}dia i Telecomunicaci\'{o},\linebreak
	Internet Interdisciplinary Institute (IN3), \linebreak
	Universitat Oberta de Catalunya (UOC),\linebreak
	Barcelona, Spain.\linebreak
	Email:\{dmegias,aqureshi\}@uoc.edu}

\begin{abstract}
Recombined fingerprints have been suggested as a convenient approach to improve the efficiency of anonymous fingerprinting for the legal distribution of copyrighted multimedia contents in P2P systems. The recombination idea is inspired by the principles of mating, recombination and heredity of the DNA sequences of living beings, but applied to binary sequences, like in genetic algorithms. However, the existing recombination-based fingerprinting systems do not provide a convenient solution for collusion resistance, since they require ``double-layer" fingerprinting codes, making the practical implementation of such systems a challenging task. In fact, collusion resistance is regarded as the most relevant requirement of a fingerprinting scheme, and the lack of any acceptable solution to this problem would
possibly deter content merchants from deploying any practical implementation of the recombination approach. In this paper, this drawback is overcome by introducing \textbf{two non-trivial improvements}, paving the way for a future real-life application of recombination-based systems. First, Nuida et al.'s collusion-resistant codes are used in segment-wise fashion \textbf{for the first time}. Second, \textbf{a novel version of the traitor-tracing algorithm} is proposed in the encrypted domain, also for the first time, making it possible to provide the buyers with security against framing. In addition, the proposed method avoids the use of public-key cryptography for the multimedia content and expensive cryptographic protocols, leading to excellent performance in terms of both computational and communication burdens. In fact, after bootstrapping, the merchant is no longer required to participate in any file transfer, reducing the investment in equipment required for content distribution to the bare minimum. The paper also analyzes the security and privacy properties of the proposed system both formally and informally, whereas the collusion resistance and the performance of the method are shown by means of experiments and simulations.
\end{abstract}


\end{frontmatter}

\section{Introduction}
\label{sec:introduction}

The reduced costs, scalability and ease of content dissemination in the digital era provide content providers with a lucrative opportunity to generate revenues through P2P systems. These benefits are the major motivation for media companies to adopt the P2P technology. However, content providers have been reluctant in choosing the P2P paradigm as a distribution vehicle to monetize digital content, since these systems are plagued with piracy. The ability to make perfect copies and the ease with which these copies can then be distributed has given rise to significant problems regarding the misuse, illegal copying and re-distribution. Copyright holders encounter uncertainties regarding the adoption or rejection of P2P networks to spread content over the Internet, since they apparently fear losing control of content distribution and worry about the promotion of illegal activity. Also, tracing a copyright violator in a P2P system with millions of connected users is an immense task. Therefore, ensuring the appropriate use of copyrighted multimedia content in P2P systems has become increasingly critical. The copyright infringement problem motivates the development of content protection techniques. Encryption can provide multimedia data with desired security during transmission by preventing them from unauthorized accessing. However, once a piece of digital content is decrypted, a dishonest customer may re-distribute it arbitrarily. Similarly, classic Digital Rights Management (DRM) systems \citep{iTunes,DRM} that are considered as a second line of defense against copyright violation, does not prove to be an effective access control against a user with the knowledge and determination to violate it. For content owners, digital watermarking proves to be a more effective anti-piracy solution. Digital watermarking has become a significant area of	research and development, and the usage of these techniques is now being considered a requisite to address the issues	faced by the proliferation of digital content. Watermarking consists of embedding a watermark, into the content that can be used later on to check the source of the content. There are two forms of watermarking, copyright watermarking and fingerprint watermarking (fingerprinting). In copyright watermarking, a watermark is embedded into the content, which indicates copyright holder’s identification. This is used to declare the copyright and cannot be used to trace the	copyright violator. On the other hand, fingerprinting techniques involve the generation of a fingerprint (a buyer-specific identification mark), the embedding operation and the development of a traitor-tracing mechanism from re-distributed copies.

In this paper, we consider two main types of participants: the merchant, who retails several items, and the buyer, who purchases (receives) an item. Traditionally, the fingerprinting scheme is usually encapsulated by running a fingerprinting protocol between the buyer and the merchant, which should prevent both parties from cheating. However, for most existing fingerprinting protocols, security comes at the cost of the buyer's privacy. In particular, in symmetric fingerprinting protocols, it is assumed that the merchants are trustworthy and always perform embedding honestly \citep{CoKiLeSh97}. However, when the buyer downloads the fingerprinted copy, the merchant can link the buyer's identity to the unique fingerprint in that copy. Apart from knowing the identity of the buyer, the merchant could also frame him/her by releasing a copy bound to him/her and accuse him/her of illegal re-distribution. Asymmetric fingerprinting schemes \citep{PfSc96} were introduced to overcome this problem. In these schemes, the fingerprint is generated through a secure dyadic protocol between the merchant and the buyer. At the end of the transaction, the merchant learns a part of the fingerprint, called the halfword, which is used to trace the buyer's identity but is not enough to frame him/her. The complete fingerprinted item is only delivered to the buyer, and thus, the merchant does not have access to the fingerprinted content obtained by the buyer. However, in both the symmetric and the asymmetric cases, the merchant still learns the identity of all buyers. This problem was solved by \citet{qn98} by introducing anonymous fingerprinting. Anonymous fingerprinting schemes retain the asymmetric property and also protect the privacy of a buyer, whose identity is only revealed and disclosed in case of illegal re-distribution. Thus, an anonymous fingerprinting protocol ensures copyright protection, privacy and security for both the buyer and the merchant simultaneously. Many fingerprinting schemes \citep{PfWa97,PfSa99,MeWo01} employed trusted third parties to provide fairness and anonymity to the merchant and the buyer, respectively. Various fingerprinting schemes do not involve trusted parties for the execution of the protocols \citep{ChSaPa03,DePr08}. Other proposals, such as \citep{Cam}, require a trusted third party only when a malicious buyer is identified and legal actions shall be taken against him/her.

Some fingerprinting protocols are based on bit-commitment schemes \citep{PfSc96,BiMe02}, which require high enciphering rates to achieve security. Other proposals, like \citep{KuTa05,PrErLa07,ARiDeBiPiPr10}, take advantage of the homomorphic property of some public-key cryptosystem. The homomorphic property allows the merchant to embed the fingerprint in the encrypted domain in such a way that only the buyer obtains the decrypted fingerprinted content. However, the use of homomorphic encryption expands data and substantially increases the communication bandwidth required for data transfers. Thus, in any case, all the proposed fingerprinting schemes involve complex cryptographic protocols, which require high bandwidth and heavy computational costs. 

A major challenge in digital fingerprinting is to thwart collusion attacks, in which several buyers, who receive copies of the same content but embedded with different fingerprints, conspire to create a new copy in which the original fingerprints are either removed or attenuated. If not properly designed, a fingerprinting system might fail to detect the traces of any fingerprints under collusion attacks with only a few colluders. Anti-collusion codes are specifically designed to counter this threat. Indeed, if the fingerprints embedded into content are codewords of anti-collusion codes, then the structure of the code allows the retrieval of the identity of at least one of the colluders. A growing number of techniques \citep{BoSh99,ChFiNaPi00,DoHe00,Tardos,SeDo03,NuFuHaKiWaOgIm07} have been proposed in the literature to provide collusion resistance in multimedia fingerprinting systems. In most of the existing fingerprinting schemes, it is assumed that the use of anti-collusion codes can make the schemes secure against collusion attacks without giving any proof of concept. 

More recently, other asymmetric fingerprinting schemes relying on $c$-secure codes have been proposed. In \citep{ChFoFuCo11}, the buyer selects the fingerprint bits from a list controlled by the merchant, but the system uses an oblivious transfer protocol which requires heavy computations. In addition, the number of communication rounds between buyer and merchant has a linear relation with the lenght of the code, making it impractical for most real-world scenarios. On the other hand, the system of \citep{Pe13} can be applied with a constant communication cost at the price of a longer codeword based on Bone-Shaw codes.

Most of the fingerprinting systems \citep{PfSc96,Cam,MeWo01,ChSaPa03,KuTa05,PrErLa07,DePr08,ARiDeBiPiPr10} found in literature are based on the traditional client-server model, where a merchant acts as a server and the buyers are the clients. Thus, the buyers obtain their fingerprinted copies of the content from the merchant in a centralized manner. This approach is non-scalable, since, for $N$ buyers, the merchant has to run the embedding algorithm $N$ times to produce $N$ fingerprinted copies. Also, the existing fingerprinting schemes involve expensive cryptographic protocols to provide secure and privacy-preserving content distribution. Often, these protocols require the presence of trusted third parties to satisfy the security and privacy requirements of anonymous fingerprinting \citep{DePr08}. Thus, these solutions are much less scalable for the merchant and involve higher CPU costs and bandwidth requirements. From the merchant's view point, the efficiency of the content distribution system is evaluated in terms of scalability and computational and communication costs. Hence, P2P distribution would be a more convenient approach for the merchant to reduce the delivery costs. However, as mentioned above, today’s P2P content distribution systems are severely abused by illegal re-distributions and the multimedia producers encounter uncertainties regarding the adoption or rejection of P2P networks to spread content over the Internet. Many systems can be found in the literature that include content protection mechanisms to solve the copyright infringement problem in P2P systems. However, a collection of identifiable personal data within P2P systems using copyright protection mechanisms raises a privacy concern among the end users. The literature review shows that very few researchers have worked on P2P content distribution systems that provide preservation of both content providers’ ownership properties and content receivers’ privacy. The next section reviews the related work on P2P content distribution systems in which either one or both copyright and privacy protection properties are satisfied.

\subsection{Previous works on P2P content distribution systems}

%In a P2P-based distribution scenario, the implementation of a fingerprinting protocol that embeds a fingerprint into the content in a secure and privacy-preserving manner, while providing collusion resistance, is a challenging task, since complex and supervised protocols are required to generate and embed the fingerprint.
The P2P content distribution systems described in the following paragraphs are designed with an intention to satisfy both copyright protection and user privacy.

\citet{slyk09} proposed an identity-based DRM system with privacy enhancement. Their DRM system retains user privacy by hiding the relationship information between users and the digital contents the users own. In order to provide strong privacy and anonymous consumption, restrictive partial blind signatures are adopted in the system. Moreover, a content key management protocol is proposed in the system to protect the users against malicious servers, and prevent them from obtaining a complete content key. However, the proposed system, in spite of employing an anonymous authentication technique to provide anonymity, does not offer traceability due to use of untraceable blind signatures.

\citet{wte11} proposed a privacy-preserving content distribution mechanism without requiring trust over any third party, by using the mechanisms of blind decryption and one-way hash chains. In the system, a privacy-preserving revocation mechanism preserves users' anonymity even if the user has been blocked for his/her misbehavior. The scheme provides privacy protection to the users by generating an anonymous token set. The proposed user only interacts with other entities in the system using these anonymous tokens. However, the proposed system does not provide any secure key management mechanism and does not support accountability parameters, i.e. it fails to discover the traitor responsible for constraints violation in the DRM system. 

\citet{DoMe13} proposed a P2P protocol for distributed multicast of fingerprinted content in which cryptographic primitives and a robust watermarking technique are used to produce different marked copies of the content for the requesting buyers. \citeauthor{DoMe13} used the reward and punishment concepts of game theory to ensure that peers rationally cooperate in P2P fashion for fingerprint embedding and content distribution. However, the system requires the implementation of a secure multi-party computation protocol between buyers for embedding the fingerprint in each transaction, resulting in increased computational and communication costs at the buyer's end. A similar idea is presented in \citep{Do11}, where the embedded information contains an expiration date used to enforce digital oblivion. 

\citet{qmr14a,qmr15} proposed a P2P content distribution framework for preserving privacy and security of the user and the merchant based on homomorphic encryption. In this framework, a few discrete wavelet transform low-frequency (approximation) coefficients are selected according to a secret key for embedding an encrypted fingerprint, exploiting the properties of a homomorphic cryptosystem. Although the selective public-key encryption of the multimedia content, which affects only a few coefficients, results in lesser data expansion, it imposes a computational burden on a merchant and an increased complexity for file reconstruction at the buyer's end. More recently, \citet{qmr16} took advantage of some of these ideas, but replaced homomorphic encryption by fragmentation, symmetric encryption, permutation and distribution of the fragments through a collection of proxy peers. Furthermore, the embedding step is pre-computed using a string of all `1's and one of all `0's and, hence, no actual embedding is required for each transaction. The proxy peers are responsible of selecting the appropriate coefficient (`0' version or `1' version) according to the specific fingerprint's bit. Permutation and symmetric encryption protect the cleartext of both the content and the fingerprint, yielding buyer frameproofness.

In \citep{MeDo13a,MeDo13}, the novel concept of fingerprinting through a recombination mechanism is proposed for a P2P-based distribution scenario. The remarkable advantage of this system is that the fingerprint embedding is required only for a few seed buyers, whereas the fingerprint of a non-seed buyer is automatically obtained as a recombination of the segments of the fingerprints of his/her source buyers. However, the traitor-tracing process requires an expensive graph search and the cooperation of some innocent buyers for tracing a copyright violator. Moreover, honest and committed proxies are required in the distribution protocol for the generation of valid fingerprints, which are encoded in two layers: the segment layer (segment-level) and the hash layer (hash-level). The segment-level encoding requires that each segment of the fingerprint is a codeword of a collusion-resistant code. In a second layer, a hash is constructed across the segments that must also be a codeword of some collusion-resistant code. The construction of a valid hash-level codeword requires the cooperation of the proxy peers.

\citet{Me14} proposed an improved version of the automatic recombined fingerprinting mechanism in which malicious proxies are considered in the distribution protocol. A four-party anonymous communication protocol is proposed to prevent malicious proxies from accessing the cleartext of the fingerprinted content. The proposed scheme provides a convenient solution for the legal distribution of multimedia contents with copyright protection while preserving the privacy of buyers, whose identities are only revealed in case of illegal re-distribution. In addition, the scheme uses standard database search for traitor tracing unlike the system in \citep{MeDo13a,MeDo13}, which requires an expensive graph search to identify an illegal re-distributor. However, the scheme proposed by \citet{Me14} still requires a two-layer anti-collusion encoding of the fingerprint, which results in a longer fingerprint and requires a complex protocol to construct valid fingerprints during P2P distribution. 

Unlike the P2P content distribution systems described in the above paragraphs, the P2P distribution systems described in the following paragraph either provide copyright protection or user privacy.

A fingerprint generation and embedding method was proposed by \citet{LKrNg10} for complex P2P file sharing networks for copyright protection. In this system, wavelet transforms and principal component analysis (PCA) techniques are used for the fingerprint generation. The wavelet technique provides a scalable approximation matrix that contains the most important low-frequency information and the PCA technique determines the orthogonal eigenvectors, which makes it possible to maximally distinguish the different fingerprints. The proposed scheme is scalable since it is able to generate a large number of unique fingerprints. The proposed framework provides a novel solution of legal content distribution, but it does not include collusion resistance or user privacy. 

\citet{hl10} proposed an asymmetric fingerprinting protocol based on a $1$-out-of-$2$ oblivious transfer protocol from the communication point of view. In the proposed scheme, multicast is exploited to reduce the bandwidth usage. The protocol uses an approach similar to the chameleon encryption, and applying  $1$-out-of-$2$ oblivious transfer, a decryption key is transmitted so that the sender does not know it. However, in the proposed fingerprinting protocol, the sender can guess the key used by the recipient with a non-negligible probability $1/n$, and the sender can even cheat in the oblivious transfer by offering the same key $n$ times, so that he/she will know the key used by the recipient. Furthermore, the proposed scheme fails to provide collusion resistance and end user's privacy. 

In \citep{ye2016}, a secure multimedia distribution framework is proposed, which combines multimedia encryption, copy detection, and digital fingerprinting to prevent widespread piracy. To deter legitimate users from illegally re-distributing the decrypted content, the scheme employs the ideas of joint copy detection and fingerprinting. Copy detection uses the content itself, rather than other information, to verify whether a protected multimedia content is a re-distributed copy or not. Fingerprinting of multimedia content using digital watermarks is an effective means of determining original owners of pirated copies by tracing down the distributor via the fingerprinting signals. Though the proposed system provides copyright protection, it fails to provide user privacy.

OneSwarm, proposed by \citet{ipka09}, is a privacy-preserving P2P system for file sharing in which the peers can query for content and download it from other users. OneSwarm provides privacy using a much different method than onion routing. The architecture of OneSwarm is based on a dense topology of peers to ensure the availability of the content, probabilistic forwarding of queries to neighbors to thwart traffic analysis, and application-level delays to thwart timing attacks. This system protects user privacy, but does not provide copyright protection.

\citet{ny15} proposed a P2P article distribution system considering both users’ privacy and criminal investigation. In the system, if many users prioritize privacy, their anonymity is protected, but if many users prioritize criminal investigation, they can trace the publisher of an article. The proposed system constructs a P2P network and spreads published articles by relaying between peers. When a peer relays an article, it records a relaying log in its local storage so that a user can trace the distribution path later on. If all users along the path from the publisher agree to cooperate in gathering their logs, they will be able to trace to the publisher. The system is not operated by a specific administrator and, hence, Pretty Good Privacy (PGP) Web of Trust (WoT) is adopted for publishing certificates. Though the proposed system protects the privacy of users from surveillance, it does not include any content protection mechanism that could protect the authors' copyrights. 

\citet{chae2016} proposed a system to prevent privacy data leakage through P2P file distribution systems. The proposed system identifies the sensitive or private data, and removes the information by the privacy data leaking risk factor from the sharing file. This process solved the issues of data loss prevention systems, such as the high detecting-error of the privacy data and invasion of privacy issues, but does not include any copyright protection mechanism. 

Table \ref{my-label} compares the P2P distribution systems in terms of adopted content protection techniques, content protection properties (copyright protection, traceability), privacy protection properties (revocable, user, data), and security against attacks (collusion/malicious, communication). In this table, a cell is marked with ``No" when no content protection technologies, or security and privacy properties, are used or guaranteed by the P2P content distribution system, respectively.

The following observations can be made with respect to the security properties:
\begin{itemize}
	\item \citet{DoMe13,qmr15,qmr16,MeDo13,Me14,LKrNg10,hl10,ye2016}, guarantee copyright protection by using fingerprinting techniques. The systems proposed by \citet{slyk09} and \citet{wte11} guarantee copyright protection by using DRM techniques. The remaining P2P content distribution systems do not offer copyright protection.
	
	\item The systems proposed by \citet{DoMe13,qmr15,qmr16,MeDo13,Me14,LKrNg10,hl10,ye2016} guarantee traceability of copyright violators by using digital fingerprinting techniques. The systems proposed by \citet{slyk09,wte11} offer copyright protection based on DRM techniques but fail to provide traceability. Similarly, the system proposed by \citet{ny15} provides traceability of a publisher by using a cooperation technique.
\end{itemize}

\begin{table}[H]
	\centering
	\caption{Comparison of P2P systems based on used security techniques and security and privacy properties}
	\label{my-label}
	\small\addtolength{\tabcolsep}{-4.0pt}
	\begin{tabular}{|l|l|c|c|c|l|c|c|c|c|c|c|c|c|}
		\hline
		\multicolumn{2}{|c|}{\multirow{3}{*}{\textbf{\begin{tabular}[c]{@{}c@{}}\\ \\P2P\\ Systems\end{tabular}}}} & \multirow{3}{*}{\textbf{\begin{tabular}[c]{@{}c@{}}\\ \\Focus\\ on:\end{tabular}}} & \multicolumn{3}{c|}{\multirow{2}{*}{\textbf{\begin{tabular}[c]{@{}c@{}}\\Content Protection \\ Mechanism\end{tabular}}}} & \multicolumn{2}{c|}{\multirow{2}{*}{\textbf{\begin{tabular}[c]{@{}c@{}}\\Security \\ Properties\end{tabular}}}} & \multicolumn{3}{c|}{\multirow{2}{*}{\textbf{\begin{tabular}[c]{@{}c@{}}\\Privacy\\ Properties\end{tabular}}}} & \multicolumn{3}{c|}{\textbf{\begin{tabular}[c]{@{}c@{}}Security against\\ Attacks\end{tabular}}} \\ \cline{12-14} 
		\multicolumn{2}{|c|}{} &  & \multicolumn{3}{c|}{} & \multicolumn{2}{c|}{} & \multicolumn{3}{c|}{} & \multicolumn{2}{c|}{\textbf{\begin{tabular}[c]{@{}c@{}}Collusion \\ Attacks\end{tabular}}} & \multirow{2}{*}{\textbf{\begin{tabular}[c]{@{}c@{}}\\Comm.\\ Attacks\end{tabular}}} \\ \cline{4-13}
		\multicolumn{2}{|c|}{} &  & \textbf{DRM} & \multicolumn{2}{c|}{\textbf{\begin{tabular}[c]{@{}c@{}}Finger-\\ printing\end{tabular}}} & \textbf{\begin{tabular}[c]{@{}c@{}}Copyright\\ Protection\end{tabular}} & \textbf{Tracebility} & \textbf{\begin{tabular}[c]{@{}c@{}}Revocable\\ Privacy\end{tabular}} & \textbf{\begin{tabular}[c]{@{}c@{}}User\\ Privacy\end{tabular}} & \textbf{\begin{tabular}[c]{@{}c@{}}Data\\ Privacy\end{tabular}} & \textbf{\begin{tabular}[c]{@{}c@{}}Content\\ Protection\\ System\end{tabular}} & \multicolumn{1}{c|}{\textbf{\begin{tabular}[c]{@{}c@{}}Privacy\\ Protection\\ System\end{tabular}}} &  \\ \hline
		\multicolumn{2}{|l|}{\begin{tabular}[c]{@{}l@{}}Sun et al.\\ (2009)\end{tabular}} & \multirow{7}{*}{\begin{tabular}[c]{@{}c@{}}\\ \\ \\ \\ \\ \\Copyright\\ and \\ Privacy\\ Protection\end{tabular}} & Yes & \multicolumn{2}{c|}{No} &Yes  & No &  No& Yes & Yes &Yes  &No  &No  \\ \cline{1-2} \cline{4-14} 
		\multicolumn{2}{|l|}{\begin{tabular}[c]{@{}l@{}}Win et al.\\ (2011)\end{tabular}} &  & Yes & \multicolumn{2}{c|}{No} & Yes & No& Yes & Yes & Yes & No & Yes & No \\ \cline{1-2} \cline{4-14} 
		\multicolumn{2}{|l|}{\begin{tabular}[c]{@{}l@{}}Domingo-\\Ferrer \&\\Meg\'{i}as\\(2013)\end{tabular}} &  & No & \multicolumn{2}{c|}{Yes} & Yes & Yes & Yes & Yes &Yes  & Yes &Yes  &No  \\ \cline{1-2} \cline{4-14} 
		\multicolumn{2}{|l|}{\begin{tabular}[c]{@{}l@{}}Qureshi\\ et al. (2015)\end{tabular}} &  & No & \multicolumn{2}{c|}{Yes} & Yes & Yes & Yes & Yes & Yes & Yes &  Yes& Yes \\ \cline{1-2} \cline{4-14} 
		\multicolumn{2}{|l|}{\begin{tabular}[c]{@{}l@{}}Qureshi\\ et al. (2016)\end{tabular}} &  & No & \multicolumn{2}{c|}{Yes} & Yes & Yes & Yes & Yes & Yes & Yes & Yes & Yes \\ \cline{1-2} \cline{4-14} 
		\multicolumn{2}{|l|}{\begin{tabular}[c]{@{}l@{}}Meg\'{i}as \&\\ Domingo-\\Ferrer\\ (2014)\end{tabular}} &  & No & \multicolumn{2}{c|}{Yes} & Yes & Yes & Yes & Yes & Yes & Yes & Yes & Yes \\ \cline{1-2} \cline{4-14} 
		\multicolumn{2}{|l|}{\begin{tabular}[c]{@{}l@{}}Meg\'{i}as\\ (2015)\end{tabular}} &  & No & \multicolumn{2}{c|}{Yes} & Yes & Yes & Yes & Yes & Yes & Yes & Yes &  Yes \\ \hline
		\multicolumn{2}{|l|}{\begin{tabular}[c]{@{}l@{}}Li et al.\\ (2010)\end{tabular}} & \multirow{3}{*}{\begin{tabular}[c]{@{}c@{}}\\ \\Copyright\\ Protection\end{tabular}} & No & \multicolumn{2}{c|}{Yes} &Yes  & Yes & No & No &Yes  &No  &No  &No  \\ \cline{1-2} \cline{4-14} 
		\multicolumn{2}{|l|}{\begin{tabular}[c]{@{}l@{}}Hu \& Li\\ (2010)\end{tabular}} &  & No & \multicolumn{2}{c|}{Yes} & Yes & Yes & No & No &Yes  &No  &No  &No  \\ \cline{1-2} \cline{4-14} 
		\multicolumn{2}{|l|}{\begin{tabular}[c]{@{}l@{}}Ye et al.\\ (2016)\end{tabular}} &  & No & \multicolumn{2}{c|}{Yes} & Yes & Yes & No & No & Yes & No & No & No \\ \hline
		\multicolumn{2}{|l|}{\begin{tabular}[c]{@{}l@{}}Isdal et al.\\ (2010)\end{tabular}} & \multirow{3}{*}{\begin{tabular}[c]{@{}c@{}}\\ \\Privacy\\ Protection\end{tabular}} & No & \multicolumn{2}{c|}{No} &No  & No & No  & Yes & Yes & No &Yes  & Yes \\ \cline{1-2} \cline{4-14} 
		\multicolumn{2}{|l|}{\begin{tabular}[c]{@{}l@{}}Tsujio \&\\Okabe\\ (2015)\end{tabular}} &  & No & \multicolumn{2}{c|}{No} & No & Yes & No & Yes & Yes & No & No & Yes \\ \cline{1-2} \cline{4-14}
		\multicolumn{2}{|l|}{\begin{tabular}[c]{@{}l@{}}Chae et al.\\ (2016)\end{tabular}} &  & No & \multicolumn{2}{c|}{No} &No  & No & No & Yes & Yes & No & No & No \\ \hline
	\end{tabular}
\end{table}

Similarly, the following observations can be made from Table \ref{my-label} in terms of privacy properties:

\begin{itemize}
	\item\citet{wte11,DoMe13,qmr15,qmr16,MeDo13,Me14} guarantee revocable privacy. In these	systems, the real identity of the users is only revealed by the trusted third party, i.e. the registration authority, in case a user is found guilty of copyright violation. The remaining systems either provide full anonymity or no anonymity at all to the users.
	\item\citet{slyk09,wte11,DoMe13,qmr15,qmr16,MeDo13,Me14,ipka09,ny15,chae2016} guarantee mutual anonymity to the users of the P2P system due to pseudonymity and anonymous authentication and communication techniques.
	\item All the compared systems guarantee data protection from unauthorized access and manipulation due to the use of symmetric/asymmetric/hybrid encryption techniques.\\
\end{itemize}

The following observations can be made with respect to security against collusion and communication attacks:
\begin{itemize}
	\item In the systems proposed by \citet{DoMe13,qmr15,qmr16,MeDo13,Me14}, the security against collusion attacks is guaranteed due to use of collusion-resistant fingerprinting. The system proposed by \citet{slyk09} offers security against collusion attacks by using a content-key management protocol. Similarly, in the P2P distribution systems proposed by \citet{wte11,DoMe13,qmr15,qmr16,MeDo13,Me14,ipka09}, the privacy of the users is preserved against malicious/collusion attacks that attempt to deanonymize the users.
	\item The communication channel used for transferring the data between two users of the systems of \cite{qmr15,qmr16,MeDo13,Me14,ipka09,ny15} are protected against malicious attacks such as man-in-the-middle attacks, denial of service and replay attacks. In the other systems, the protection against communication attacks is either not provided or not discussed.
\end{itemize}

It is apparent, from Table \ref{my-label}, that most of the presented P2P distribution systems focus on either providing copyright protection to content owners or privacy to end users. Only a few P2P content distribution systems \citep{DoMe13,MeDo13,Me14,qmr15,qmr16} exist that focus on \textbf{both} copyright protection (with collusion resistance) and user privacy and anonymity. For this reason, these are the systems selected for a detailed comparison with the proposed scheme in Section \ref{sec:comparative}.

\subsection{Contribution and plan of the paper} 
The main contribution of this paper is to introduce a new fingerprinting protocol based on recombination approach that provides a secure, anonymous and efficient collusion-resistant-based asymmetric fingerprinting scheme within a P2P content distribution system. Following are the properties of our contribution:

\begin{enumerate}
	\item In order to provide collusion resistance against $c$ colluders, state-of-the-art collusion-resistant fingerprinting codes (\citet{NuFuHaKiWaOgIm07} codes) are employed, which provide small length fingerprint codewords. \textbf{To the best of our knowledge, these codes are used segment-wise for the first time in our proposal}. The segment-wise construction of the fingerprint using \citeauthor{NuFuHaKiWaOgIm07}'s codewords prevents the two-layer encoding required by the previous recombination-based proposals \citep{MeDo13,Me14} and simplifies the distribution protocol to a great extent. Unlike \citep{MeDo13,Me14}, the proposed scheme does not require the cooperation of the proxy peers to construct a valid hash-level code to be used for colluder(s) tracing. Hence, the fingerprint generated by recombination is always valid and does not require further verification, which is a clear advantage compared to the schemes proposed in \citep{MeDo13a,MeDo13,Me14}.
	\item The proposed scheme provides traceability for the identification of the colluder(s). \textbf{The standard \citeauthor{NuFuHaKiWaOgIm07}'s traitor-tracing algorithm has been redesigned to perform traitor tracing in the encrypted domain} such that buyer frameproofness is retained. We believe that this is the first time that the traitor-tracing algorithm is performed in an encrypted domain. The scores of the colluder(s) calculated using the proposed tracing protocol match those that would be calculated in the cleartext domain.
	\item The use of \citeauthor{NuFuHaKiWaOgIm07}'s codewords to construct fingerprints in a segment-wise form is analysed and a threshold to identify the colluders (with no false positives or negatives) in the traitor-tracing protocol is derived. Simulations with the proposed traitor-tracing system (with collusion) are provided to show the resistance of the fingerprinting scheme against various collusion attacks.
	\item To perform efficiently in P2P network, differing from the existing fingerprinting schemes \citep{qmr14a,qmr15}, heavy-burden operations, such as public-key cryptography, are applied only for data signing and for the encryption of short bit strings, such as the binary fingerprints. In the distribution protocol, symmetric cryptography is used to encrypt the fragments of the content in order to provide data security.
	\item Formal and informal proofs are made available to show that the system exhibits security and revocable privacy to the content owners and the buyers, respectively.
\end{enumerate}

The rest of this paper is organized as follows. Section \ref{sec:previous} reviews the main features of the proposals suggested in \citep{MeDo13a,MeDo13} and \citep{Me14}, and also highlights the main drawbacks of these systems. Section \ref{sec:proposed} details the protocols and algorithms of the proposed system in order to overcome the drawbacks of the mentioned schemes. Section \ref{sec:Features} discusses the features of the new system and compares it with existing anonymous fingerprinting proposals for P2P distribution networks. Section \ref{sec:theoretical} analyses the security and privacy of the proposed system. This section also presents simulated and theoretical results designed to evaluate the collusion resistance of the new fingerprinting scheme. Finally, Section \ref{sec:conclusions} summarizes the conclusions and future research issues.

%%%%%%%%%%%%%%%%%2. Related Work %%%%%%%%%%%% % % % % % % % % % % % % % % % % % % % % % % % % % % % % % % % % % %

\section{Related work}
\label{sec:previous}
The fingerprinting system presented in this paper stems from the work proposed in \citep{MeDo13a,MeDo13} and \citep{Me14}, which introduced the concept of automatically recombined fingerprints for P2P networks. In the following sections, we present the main features and drawbacks of these systems.
% % % % % % % % % % % % % % % % % % % % % % % % % % % % % % % % % % % % % % % % % % % % % % % % % % % % % % % % % %

% % % % % % % % % % % % % % % % % % % % % % % % % % % % % % % % % % % % % % % % % % % % % % % % % % % % % % % % % %
\subsection{Features of the system proposed by \protect\citet{MeDo13a,MeDo13}}
\label{sec:features1}
The main features of the fingerprinting scheme proposed in \citep{MeDo13a,MeDo13} are the following:
\begin{itemize}
	\item The system consists of a merchant, seed buyers, non-seed buyers, a group of P2P proxies, a transaction monitor and a tracing authority. The proxy (or a set of proxies) and the tracing authority are considered as trusted parties, whereas the remaining entities are not trusted.
	\item The content is divided into several ordered fragments and each of them is embedded separately with a  binary sequence.
	\item The binary sequence embedded in each fragment is called a ``segment" and the concatenation of all the segments forms the binary fingerprint.
	\item The merchant creates a set of $M$ seed copies and distributes them to the $M$ seed buyers. The fingerprints are generated by the merchant in such a way that they have low pair-wise correlation.
	\item A non-seed buyer obtains the fragments of the content from at least two other buyers and his/her fingerprint is built as a recombination of the segments of his/her parents as shown in Figure \ref{fig:fig1}.
	\item The embedding process is required only for $M$ seed buyers. For the non-seed buyers of the system, different fingerprints are created without any further execution of the embedding scheme.
	\item To ensure anonymous communication between buyers, an onion routing-like protocol using a P2P proxy (or a chain of proxies) is used within the system.
	\item The proxies do not know the real identities, only pseudonyms, of the source and destination buyers. In the anonymous data transfer protocol, the proxies are responsible for forwarding a one-time symmetric session key from the child buyer to the parent buyer in order to encrypt the fragments of the content for data security.
	\item The proxies register each transaction at the transaction monitor and also inform it about the number of fragments transferred to the buyer. For each fragment, the proxies receive a single-bit hash of the corresponding segment. Each proxy forms a bit string with the hash bits of all the transferred segments and encrypts this string using the public key of the transaction monitor. When the child buyer receives all the fragments of the complete content, the proxies share their encrypted strings, concatenate them, and encrypt them again, one time per each parent buyer, using the public-key of the parent buyer, creating the encrypted version of the hash. Thus, no proxy has access to the complete cleartext of the fingerprint's hash. The transaction monitor contacts all the corresponding proxies to request the encrypted hash and stores it in its transaction database (one time per parent buyer).
	\item The transaction monitor keeps a record of each transaction in order to keep track of data transfer between the buyers. These records contain the identifier of the purchased content (e.g. a perceptual hash), the pseudonyms of the involved buyers (source and destination) and the encrypted hash of the fingerprint. The hash of the fingerprint is encrypted using the public key of the transaction monitor and the public key of each parent buyer. In the execution of the traitor-tracing protocol, the private key of at least one parent buyer is required by the transaction monitor to decrypt the encrypted hash.
	\item The merchant has access to the buyers' database, thus only he/she knows the true identities of the seed and non-seed buyers of the system.
	\item In case an illegally re-distributed copy is found, the re-distributor is identified through a graph search directed by a binary correlation function between the fingerprint extracted from the re-distributed copy and the fingerprints of the tested buyers. Cooperation is required from some honest buyers during the search process. The fingerprints' hashes stored in the transaction monitor are enough to prevent those buyers from cheating in this step.
	
	%%%%%%%%%%%%%%%Figure 1%%%%%%%%%%%%%%%%%%%%%%%
	\begin{figure}[ht]
		\centering
		%\graphicspath{G:\PhD Docs\elsarticle-ecrc\heredity.pdf}
		\includegraphics[width=0.7\textwidth]{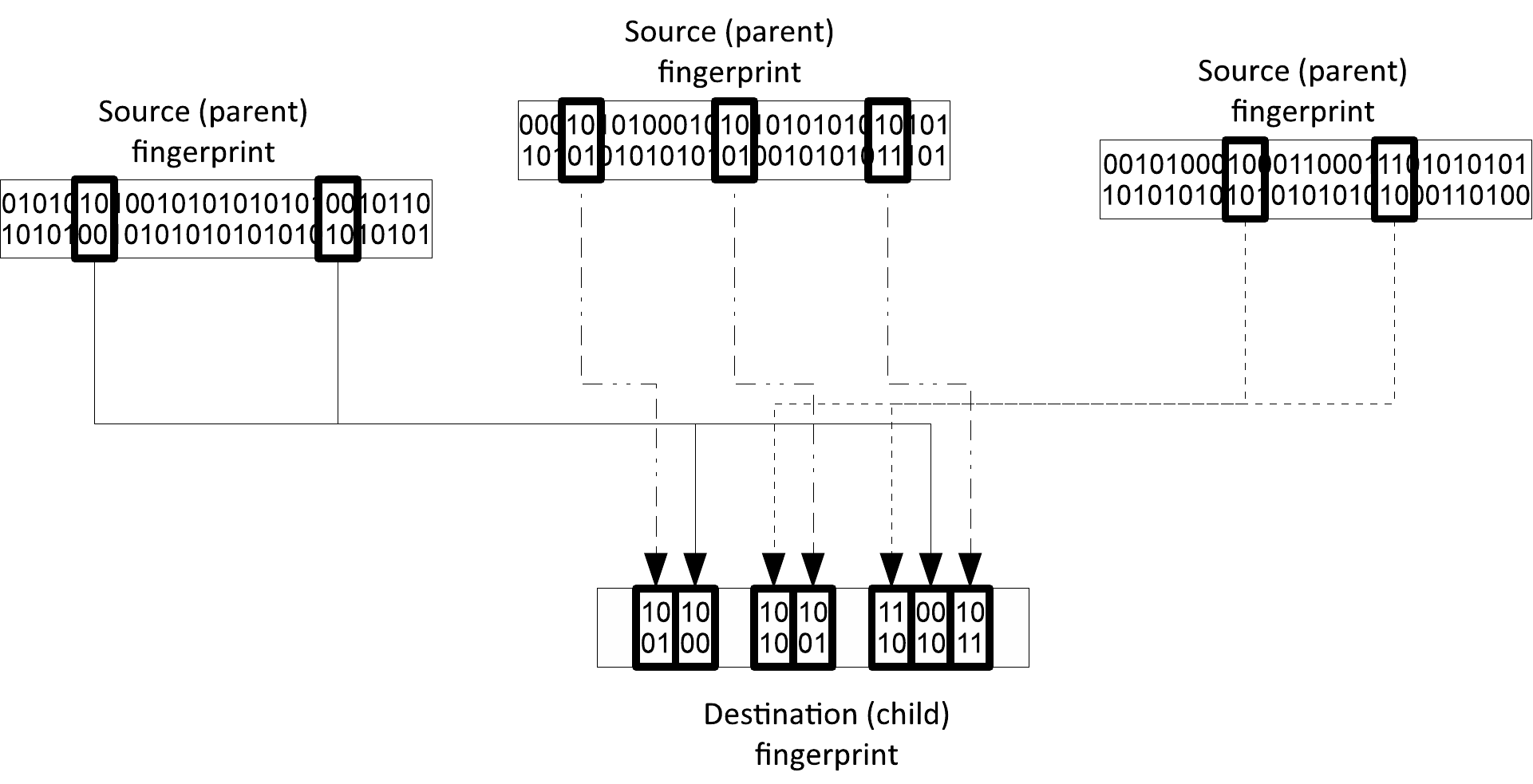}
		\caption{Automatic recombined fingerprint construction}
		\label{fig:fig1}
	\end{figure}
	%%%%%%%%%%%%%%%%%%%%%%%%%%%%%%%%%%%%%%%%%%%
	
	\item In each step of the traitor-tracing protocol, the buyer with the maximum correlation is chosen as the most likely ancestor of the illegal re-distributor. Occasionally, a higher correlation may be obtained for a non-ancestor of the buyer of the illegally re-distributed copy, which would require the exhaustion of a sub-graph and backtracking.  
	\item The search ends when perfect correlation is found between the traced fingerprint and the fingerprint of the tested buyer. In case a buyer refuses to cooperate with the tracing authority in the tracing-tracing protocol, the fingerprint hash recorded by the transaction monitor can be used as an evidence against him/her.
	\item Anti-collusion codes are used in the protocol to provide collusion resistance against attackers. Each segment is encoded with an anti-collusion code, which can be used to reconstruct the segment of at least one of the colluders.
	\item The fingerprints are constructed in such a way that their hashes are also codewords of some collusion-resistant code. Thus, in the traitor-tracing protocol, once the segments are reconstructed by the tracing authority, the hash of at least one of the colluders can be obtained. However, during the distribution protocol, the fragments must be distributed by the proxies following some (non-specified) protocol, with the participation of the transaction monitor, to construct a valid anti-collusion codeword for each hash. 
	\item To obtain a segment with the required hash bit, the proxies contact the potential parents subsequently, since only the parents having that particular hash bit would be accepted as the source for the specific fragment of the content. In this way, a valid codeword is formed for the hash.
	\item In case of collusion, in order to identify an illegal re-distributor in the traitor-tracing protocol, the correlation function is applied with the hash of the traced fingerprint and that of a tested buyer as inputs, instead of using the fingerprints. 
\end{itemize}

In comparison to previous fingerprinting schemes \citep{PfSc96,Cam,KuTa05,ARiDeBiPiPr10}, which require highly demanding technologies such as public-key encryption of the contents, secure multi-party protocols or zero knowledge proofs, the system proposed in \citep{MeDo13a, MeDo13} offers remarkable advantages to both buyers and the merchant in terms of computational and communication costs: 1) the embedding is required only for a few seed buyers ($M$) and the fingerprint of the other buyers are automatically generated as a recombination of segments; 2) the fingerprints of non-seed buyers are unknown to the merchant (achieving buyer frameproofness); 3) transactions between buyers are fully anonymous; 4) public-key cryptography is restricted to the transmission of short-bit strings; and 5) an illegal re-distributor can be identified through the traitor-tracing protocol.

\subsection{Drawbacks of the system proposed by \protect\citet{MeDo13a,MeDo13}}
\label{sec:drawbacksMeDo}
Despite the advantages of this scheme, detailed in Section \ref{sec:features1}, it exhibits the following shortcomings:
\begin{enumerate}
	\item The traitor-tracing process requires an expensive graph search in order to identify an illegal re-distributor.
	\item Some innocent buyers are requested to cooperate with the tracing authority in the traitor-tracing protocol. This participation can result in the following problems:
	\begin{itemize}
		\item The number of tested buyers in the traitor-tracing process is not known a priori. Thus, in a worst-case situation, the whole set of the buyers would have to be checked in order to find the illegal re-distributor, resulting in increased computational and communication costs. 
		
		As remarked above, the graph search carried out for traitor tracing my require backtracking, as shown in a graphical example in Figure \ref{fig:fig3}, where buyer $B_{48}$ is the illegal re-distributor of the content. The graph search jumps from buyer $B_4$ to buyer $B_{51}$ skipping the illegal re-distributor ($B_{48}$). Once the sub-graph beyond $B_{51}$ is exhausted, the traitor-tracing system backtracks and goes back to buyer $B_{17}$, as shown by a dotted arrow, before finally reaching the illegal re-distributor after two more hops. 
		
		%%%%%%%%%%%%%%%Figure 3%%%%%%%%%%%%%%%%%%%%%%%
		\begin{figure}[ht]
			\centering
			%\graphicspath{G:\PhD Docs\elsarticle-ecrc\tracing1.pdf}
			\includegraphics[width=6cm]{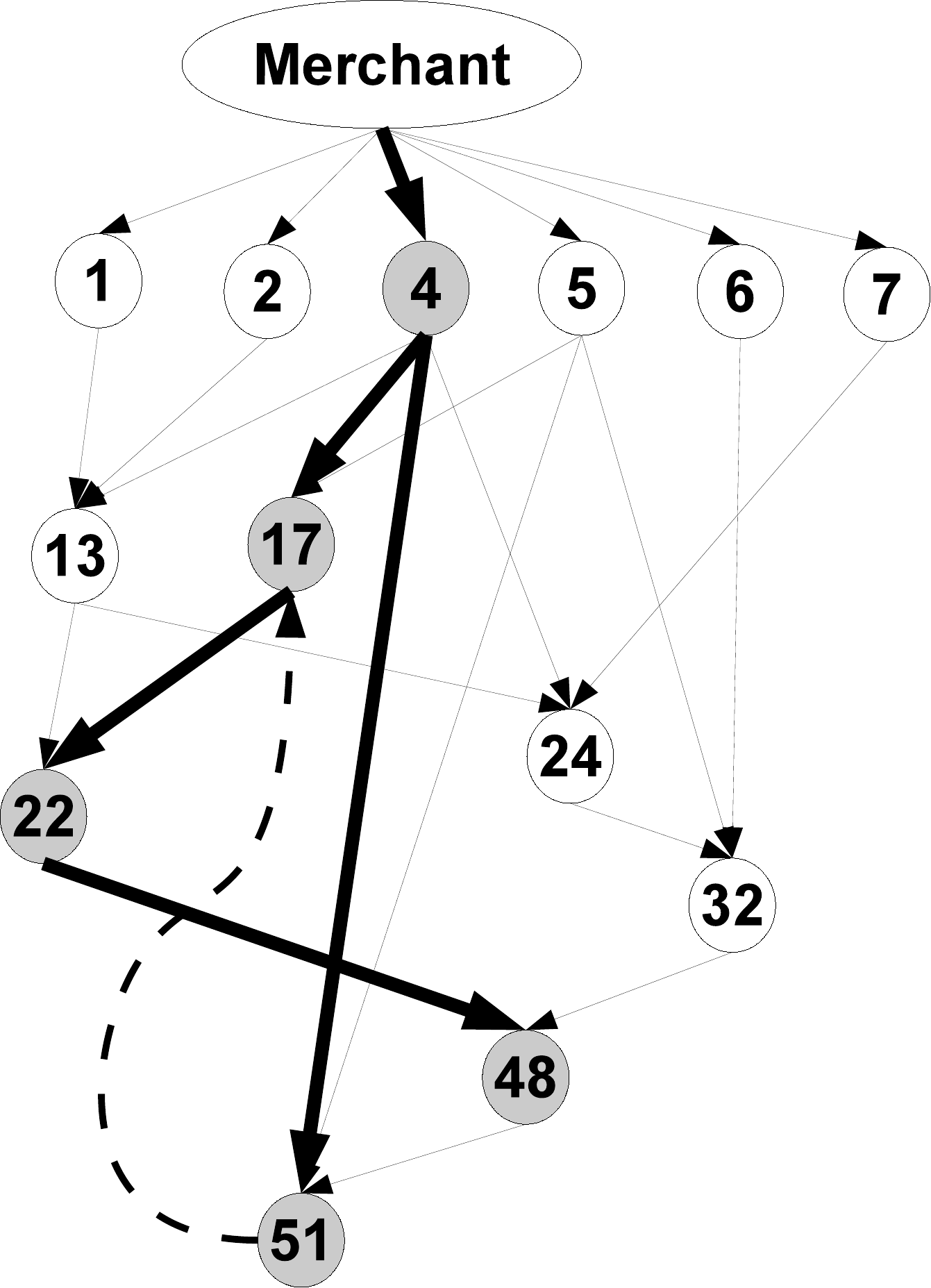}
			\caption{Traitor-tracing graph search protocol}
			\label{fig:fig3}
		\end{figure}
		%%%%%%%%%%%%%%%%%%%%%%%%%%%%%%%%%%%%%%%%%%%
		
		\item In some distribution scenarios, if a buyer is unable to participate in the traitor-tracing protocol either due to a genuine problem or a malicious reason, it may result in non-traceability of an illegal re-distributor.
		\item The privacy of an honest buyer is surrendered to some extent, since the tracing authority can associate the purchased content with the buyer's sensitive information such as his/her IP address.
		\item The fingerprint of the buyer does not remain secret, since the tracing authority needs to compute the correlation between the buyer's and the traced fingerprint. Thus, in case a malicious buyer re-distributes the content illegally, he/she can argue that his/her fingerprint could have been leaked during the traitor-tracing process of a former illegal re-distribution.
	\end{itemize}
	\item The distribution protocol relies on honest proxies for data transfer between the buyers of the system. This trust on the proxies is detrimental to the privacy interests of the buyers, since the proxies could act maliciously and use the session keys transmitted from the parent buyer to the child buyer so as to decrypt the transferred encrypted content. Consequently, a collusion of all malicious proxies is enough to obtain the fingerprinted content and frame an honest buyer for illegal re-distribution.
	\item The anti-collusion mechanism used in this system has several drawbacks that are shared with its improved version \cite{Me14}, as detailed in Section \ref{sec:drawbacksMe14}.
\end{enumerate}

% % % % % % % % % % % % % % % % % % % % % % % % % % % % % % % % % % % % % % % % % % % % % % % % % % % % % % % % %

\subsection{Features of the system proposed by \protect\citet{Me14}}
The use of automatic recombined fingerprints proposed in \citep{MeDo13a,MeDo13} allows the multimedia producers and merchants to save bandwidth and CPU time thanks to the P2P distribution of the content and the automatic generation of fingerprints through recombination. In addition, the system provides buyers with privacy preservation and buyer frameproofness. The traceability of a source responsible for an illegal re-distribution of the content is also possible. However, the aforementioned system has some shortcomings that can make it inefficient and impractical under certain circumstances, as detailed in Section \ref{sec:drawbacksMeDo}. 

To overcome the major drawbacks of \citep{MeDo13a,MeDo13}, \citet{Me14} proposed a more efficient and practical content distribution system using recombined fingerprints as its main building block. This modified system inherits some features from the scheme described in \citep{MeDo13a,MeDo13}. Thus, only the novel and improved features of the referred method are highlighted below:
\begin{itemize}
	\item  Proxies are not trusted and the security against malicious proxies is obtained through a four-party anonymous communication protocol that prevents proxies from accessing the cleartext of the fragments of the content.
	\item In addition to the pseudonyms of the involved buyers, the perceptual content hash, the encrypted hash of the fingerprint, and the transaction date and time, the transaction monitor also stores the fingerprints of the buyers in encrypted form.
	\item The hash of the fingerprint is encrypted only with the public key of the transaction monitor ($K_c$).
	\item Each fragment of the content is transmitted, with a fingerprint's segment $g_j$ embedded into it,  together with an encrypted version of the segment $E(g_j,K_c)$.
	\item Each proxy selects a set of $m$ contiguous fragments of the content that carry $m$ contiguous segments of the fingerprint embedded into them. The proxies also store the corresponding encrypted segments, $E(g_j,K_c)$. The construction of the fingerprint with segments and sets of contiguous fragments is illustrated in Figure \ref{fig:fig2}. The number of segments $n_s$ does not need to be a multiple of $m$, since a larger set in can be arranged for the last set of segments of the fingerprint, as shown in the figure. For example, if $m=10$ and the total number of segments is $n_s=74$, the last set of segments ($s_L$) would be formed by 14 segments instead of 10. In any case, we have $0\leq n_s-Lm<m$.
	
	%\begin{figure}[ht]
	%\centering
	%\graphicspath{G:\PhD Docs\elsarticle-ecrc\Fingerprint_david.pdf}
	%\includegraphics[width=15cm,height=2cm]{Fingerprint_david.pdf}
	%\caption{Fingerprint's segments ($g_j$) and sets of $m$ contiguous fragments}
	%\label{fig:fig2}
	%\end{figure}
	%%%%%%%%%%%%%%%Figure 2%%%%%%%%%%%%%%%%%%%%%%%
	\begin{figure}[ht]
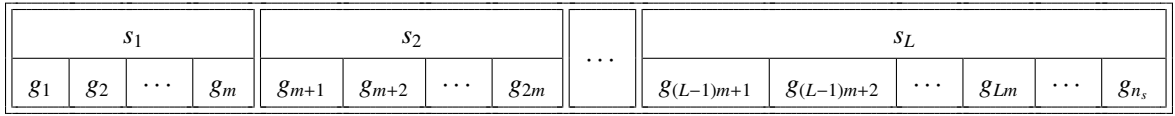

		\centering
		%\begin{center}
		\begin{tabular}{||c|c|c|c||c|c|c|c||c||c|c|c|c|c|c||}
			\hhline{|t:=:=:=:=:t:=:=:=:=:t:=:t:=:=:=:=:=:=:t|}
			\multicolumn{4}{||c||}{$s_1$} & \multicolumn{4}{c||}{$s_2$}  & \multicolumn{1}{c||}{\multirow{2}{*}{$\cdots$}} & \multicolumn{6}{c||}{$s_L$} \\
			\hhline{||----||----||~||------||}
			$g_1$ & $g_2$ & $\cdots$ & $g_{m}$& $g_{m+1}$ & $g_{m+2}$ & $\cdots$ & $g_{2m}$ &
			& $g_{(L-1)m+1}$ & $g_{(L-1)m+2}$ & $\cdots$ & $g_{Lm}$ & $\cdots$ & $g_{n_s}$ \\
			\hhline{|b:=:=:=:=:b:=:=:=:=:b:=:b:=:=:=:=:=:=:b|}
		\end{tabular}
		\caption{Fingerprint's segments ($g_j$)  and sets ($s_k$) of $m$ contiguous segments}
		\label{fig:fig2}
		%\end{center}
	\end{figure}
	%%%%%%%%%%%%%%%%%%%%%%%%%%%%%%%%%%%%%%%%%%%
	\item The proxy encrypts the concatenation of $m$ contiguous encrypted segments using the public key of the tracing authority ($K_a$) and sends it to the transaction monitor. The possibility of a proxy accessing the complete sequence of the encrypted segments is unrealisable due to the fact that each buyer is required to choose at least two proxies.
	\item The transaction monitor serves as a temporary database for storing the session keys shared between the parent and child buyers in the anonymous data transfer protocol. A pseudorandom number $r$ is generated by a parent buyer and is used as a handle for the shared key $K$, which is sent by the parent to the transaction monitor. During data transfer, the parent buyer sents $r$ to the child buyer through the proxy. On receiving $r$, the child buyer sends it to the transaction monitor, who replies with the corresponding session key $K$. The corresponding register is blocked and, after a certain time, it is finally removed by the transaction monitor.
	\item A malicious proxy who tried to access the database to retrieve $K$ in order to obtain the cleartext of the fragments of the transferred content would be detectable, since the register containing $K$ would be either blocked or removed by the transaction monitor (either for the proxy or for the child buyer).
	\item In case an illegally re-distributed copy is found, the tracing authority extracts the fingerprint from it and uses the public key of the transaction monitor and its own public key to produce the encrypted fingerprint. This encrypted fingerprint is searched in the database of the transaction monitor to disclose the pseudonym of the illegal re-distributor.
	\item In case of collusion, the hash of at least a colluder can be reconstructed from the colluded fingerprint. This hash is encrypted by the transaction monitor and searched in the transaction database to retrieve the pseudonym of the colluder.
	\item Finally, the merchant reveals the real identity of the copyright violator by searching the pseudonym provided by the transaction monitor in his/her database of buyers. 
\end{itemize} 
% % % % % % % % % % % % % % % % % % % % % % % % % % % % % % % % % % % % % % % % % % % % % % % % % % % % % % % % % % %

% % % % % % % % % % % % % % % % % % % % % % % % % % % % % % % % % % % % % % % % % % % % % % % % % % % % % % % % % % % %
\subsection{Drawbacks of the system proposed by \protect\citet{Me14}}
\label{sec:drawbacksMe14}
\citet{Me14} provided several significant improvements in the fingerprinting scheme of \citep{MeDo13a,MeDo13} to overcome the major drawbacks of the recombination-based fingerprinting approach, such as a simpler traitor-tracing algorithm, which does not require cooperation of honest buyers for tracing a copyright violator, and a four-party anonymous communication protocol that prevents malicious proxies from framing an innocent buyer for illegal re-distribution. Though the system satisfies all the necessary security and privacy requirements of an anonymous fingerprinting scheme, it still exhibits the following drawbacks:
\begin{enumerate}
	\item The fingerprinting scheme requires a two-layer anti-collusion code (segment level and fingerprint level), which results in a longer codeword. 
	\item As in the original scheme \citep{MeDo13a,MeDo13}, the system in \citep{Me14} requires the hash of the fingerprint to trace an illegal re-distributor in case of collusion of several buyers. To achieve this goal,
	the hash of the fingerprint must be a valid codeword of some anti-collusion code. Thus, the monitor requires the cooperation of the proxies to construct a valid hash-level codeword. This demands a communication channel between the P2P proxies who are carrying the content from at least two parent (source) buyers to the child (destination) buyer. These proxy peers must communicate with each other in the presence of the transaction monitor, who must verify that the constructed fingerprint hash is valid, by indicating the correct hash bits for specific segments. Hence, a  complex (non-specified) protocol involving all proxies (of the same file transfer) and the transaction monitor is needed to generate a valid fingerprint. This protocol must be secure enough to guarantee that the proxies cannot collude to disclose the whole hash-level codeword, which may compromise an innocent buyer.
\end{enumerate}

\citet{Me14} did not address these problems, both of them related to the two-layer anti-collusion coding of the fingerprint. As shown below, this drawback can be overcome by introducing state-of-the-art collusion-resistant codes only at segment level (the requirement of a hash-level codeword can be removed) combined with a less complex traitor-tracing protocol. Thus, the system proposed in this paper is designed to remove these drawbacks and improve the collusion resistance mechanism of recombined fingerprinting. 

% % % % % % % %Section 3. Proposed System % % % % % % % % % % % % % % % % % % % % % % % % % % % % % % % % % % % % %
\section{Proposed system}
\label{sec:proposed}
This section describes the novel system, which is designed to retain the basic idea of automatically recombined fingerprints proposed in \citep{MeDo13a,MeDo13,Me14} and solves the major problems of the  system of \citep{Me14}, which are discussed in Section \ref{sec:drawbacksMe14}.
% % % % % % % % % % % % % % % % % % % % % % % % % % % % % % % % % % % % % % % % % % % % % % % % % % % % % % % %
\subsection{System entities}
\label{sec:entities}
The proposed system involves six entities and the function of each entity is defined as follows:
\begin{itemize}
	\item\textbf{Merchant:} A merchant (\textit{Me}) is an entity that has access to the database of the buyers. Thus, he/she is the only party who knows the true identities of the buyers of the system. It also receives the payments associated to the purchases of the buyers. 
	\item\textbf{Seed buyers:} The seed buyers ($B_{i}$ for $i=1,\ldots, M$) are dummy buyers of the system. These buyers receive the fingerprinted copies of the content from the transaction monitor that are used by the system to bootstrap the P2P distribution protocol. The seed buyers are contacted by the buyers of other generations to obtain further copies of the content. 
	\item\textbf{Other buyers:} The other buyers ($B_{i}$ for $i=M+1,\ldots, N$, with $M\ll N$) obtain their fingerprinted copies through the P2P distribution system such that the binary fingerprint of their copy is a binary sequence automatically formed as the recombination of the segments of the embedded fingerprints of their parents (source buyers). Once a buyer obtains the content, he/she can become the source of the content for other buyers.
	\item\textbf{Proxies:} A proxy is an entity that is used to create an anonymous connection between the buyers such that each source and destination buyer does not loose his/her anonymity. They provide anonymous communication between buyers through a specific protocol similar to an onion routing mechanism, based on Chaum's mix networks \citep{Ch81}. The proxies are also responsible for selecting at least two buyers (parents) as the sources of the content fragments for a destination buyer in the anonymous P2P distribution protocol.
	\item\textbf{Transaction monitor:} The transaction monitor (\textit{MO}) is an entity that generates collusion-resistant fingerprint codes for the $M$ seed buyers and embeds these codes into the fragments of the multimedia content in order to produce $M$ fingerprinted seed copies. \textit{MO} is also responsible for maintaining a transaction database that keeps record of the permuted and encrypted fingerprints embedded in the purchased contents along with the pseudonyms of the buyers involved in the transactions.
	\item\textbf{Tracing authority:} In our system, the tracing of illegal re-distributors is carried out by a tracing authority ($T_A$) in cooperation with \textit{MO}. This is due to the fact that the presence of a single party responsible for both fingerprint generation and tracing of an illegal re-distributor could result in a security compromise of a buyer, if the foregoing entity unexpectedly behaves in a malicious manner. Since the database of the buyers' fingerprints is maintained by \textit{MO}, it could falsely accuse or frame an honest buyer either for an illegal re-distribution of the content (piracy) or forming a coalition with other malicious party (collusion). In the proposed system, $T_{A}$ is responsible for tracing  illegal re-distributors by executing a relevant part of the traitor-tracing protocol in case of piracy claim. As detailed below, this traitor tracing is carried out by $T_A$ in conjunction with \textit{MO}.
\end{itemize}
% % % % % % % % % % % % % % % % % % % % % % % % % % % % % % % 

\subsection{Design requirements and assumptions}
\label{sec:design}
In this section, the design requirements and security assumptions of the system are described.
\subsubsection{Design requirements}
\label{sec:requirements}
In the following, the design requirements related to the construction of the proposed system are defined:
\begin{itemize}
	\item The fingerprinting scheme should be collusion resistant against a given maximum number of colluders ($c\leq c_0$) as specified by \citet{NuFuHaKiWaOgIm07} codes.
	\item At the time of registration, each buyer can authenticate himself/herself to \textit{Me} and obtain a pseudonym for use within the system. Once a pseudonym is obtained, each buyer should sign a document (\textit{AGR}) using his/her private key, that certifies the association between the pseudonym and his/her real identity. 
	\item The real identity of $B_{i}$ should remain anonymous during transactions unless he/she is proven guilty of copyright violation. 
	\item The possible collusion of malicious proxies should be unable to frame an honest buyer. Similarly, any malicious party should not be able to construct the fingerprinted copy corresponding to any buyer to frame an honest user of the system.
	\item The identity of $B_{i}$ should not be linked to his/her activities such as purchasing, transferring of content and so on.
	\item \textit{MO} and the proxies should not have access to either the real identities of the buyers or the cleartext of the fingerprints.
	\item Each buyer must choose at least two proxies (or chains of proxies) in the P2P distribution protocol.
	\item The communication between the peer buyers within the P2P distribution system must be anonymous in order to preserve the buyers' privacy.
	\item The data transferred through proxies should be encrypted in such a way that only the sender and the receiver have access to their cleartext.
	\item In case \textit{Me} finds a pirated copy, $T_A$ should be able to trace and identify an illegal re-distributor with the help of \textit{MO}. 
	\item In the traitor-tracing protocol, $T_A$, with the help of \textit{MO}, should be able to identify the illegal re-distributor without requiring the cleartext of the fingerprint of honest buyers, i.e. the traitor-tracing algorithm of \citet{NuFuHaKiWaOgIm07} codes should be executed in the encrypted domain. 
\end{itemize}

\subsubsection{Security assumptions}
\label{sec:assumptions}
The security assumptions of the proposed system are the following:
\begin{itemize}
	\item \textit{Me} is not trusted either for content distribution or for the association of pseudonyms with the corresponding real identities of the buyers. Hence, proofs (\textit{AGR}) shall be provided by the merchant to authenticate the provided information.
	\item The $M$ seed buyers are fake buyers or dummies that serve as special-purpose servers for the other generation ($M+1,\ldots,N$) buyers. The seed buyers will not participate in any collusion and thus, their fingerprints are never used in the traitor-tracing protocol. \textit{MO} is not required to keep a record of the fingerprints corresponding to the $M$ seed buyers.
	\item The $N-M$ non-seed buyers $B_i$ (for $i=M+1,\ldots,N$) are not  trusted entities of the system and the protocols are designed to guarantee that they are transferring authenticated fragments of the content when they become sources of the content for other buyers.
	\textit{MO} is trusted for managing the session key database and the transaction database used in the anonymous P2P content distribution protocol. The details are provided in the description of Protocols \ref{protocol1} and \ref{protocol2}.
	\item \textit{MO} reveals the true pseudonym corresponding to one or more illegal re-distributors or  colluders in the traitor-tracing protocol.
	\item Each binary fingerprint $f_i$ of a seed buyer, generated by \textit{MO}, is composed of multiple ordered segments $g_j$, where each segment is generated by using \citet{NuFuHaKiWaOgIm07} $c$-secure fingerprinting codes. The length of the fingerprint $l$ is a known constant of the system.
	\item $T_A$ is a trusted entity. Hence, it is expected that $T_A$ does not form a coalition with any other party to frame an innocent buyer or break anyone's privacy. 
	\item The length of a permutation key $\sigma$, generated by $T_A$, is equal to the size of the fingerprint, i.e. $l$. The purpose of $\sigma$ is to prevent \textit{MO} from disclosing the fingerprint of any innocent buyer through the inspection of the encrypted fingerprints stored in his/her database. $\sigma$ is kept secret from \textit{Me}, \textit{MO} and the proxies for the security of the stored fingerprints.
	\item The fragments of the content transferred through the proxies in the anonymous distribution protocol are encrypted using symmetric cryptography.
	\item The watermarking method used for embedding the fingerprint $f_i$ in the seed copies is imperceptible, secure and robust against common signal processing attacks. 
	\item Each entity (\textit{Me}, \textit{MO}, $B_{i}$, $T_A$, proxies) is supposed to have a public key $K_p$ and a private key $K_s$. 
	\item The public keys must be authenticated so that a party who uses these keys knows that these belong to the legitimate parties. To achieve this, the existence of a PKI is assumed.
\end{itemize}

% % % % % % % % % % % % % % % % % % % % % % % % % % % % % % % % % % % % % % % % % % % % % % % % % % % % % % % % % %

\subsection{Attack model}
\label{sec:model}
This section describes the main attacks that may be performed on the protocols of the proposed system. These attacks can be aimed to break either the security or the privacy properties of the system. In order to design a secure and privacy-preserving distribution system, the following three properties shall be guaranteed:
\begin{itemize}
	\item \textbf{\textit{Copyright protection:}} This property requires that a non-seed buyer accused of illegally re-distributing a content should not be able to repudiate his/her guilt of copyright violation.
	\item \textbf{\textit{Buyer frameproofness:}} It requires that any malicious party of the system or a possible collusion of such parties should not be able to accuse an innocent buyer of illegal re-distribution.
	\item\textbf{\textit{Buyer's privacy:}} The identity of a buyer should remain anonymous during transactions and it should not be linked to his/her online activities.
\end{itemize} 
In the following, we describe an attack model for the proposed system related to the buyer's privacy and security attacks from malicious entities and the resistance of a fingerprint against collusion attacks.
\subsubsection{Framing attacks}
\label{sec:framing}
The following type of attacks are aimed to accuse (frame) an innocent buyer of illegal re-distribution of the purchased content. 
\begin{enumerate}
	%\item\textbf{\textit{Buyer's rights problem:}} The merchant has an access to all the fragments of $M$ fingerprints received from $T_A$ and he/she may recombine the fragments of these fingerprints to produce a new fingerprinted copy of the content. In case this copy is re-distributed, the malicious merchant may accuse an innocent buyer of illegal re-distribution.
	\item\textbf{\textit{Proxy authentication attacks:}} An attacker may impersonate a proxy in the system and try to obtain fragments of a content from different buyers in order to create a fake colluded copy of the content. Consequently, in the traitor-tracing algorithm, one or more honest buyers may be identified as traitors, since these buyers provided their fragments to the fake proxy in the distribution protocol.
	\item\textbf{\textit{Man-in-the-middle attack:}} An attacker may attempt to eavesdrop on a communication between a buyer and one or more of his/her proxies and store all the fragments of the content.
	\item\textbf{\textit{Database attack:}} An attacker may try to obtain the fingerprint of a buyer stored in \textit{MO}'s database.
	\item\textbf{\textit{Conspiracy attacks:}} A possible coalition of one or more of the participants (proxies, \textit{Me}, $B_i$, \textit{MO} or $T_A$) in the distribution and traitor-tracing protocols may try to obtain the fingerprint or the fingerprinted content linked to a particular honest buyer.
	\item\textbf{\textit{Ciphertext only attacks:}} \textit{MO} may try to disclose the fingerprint of an honest buyer through inspection of the encrypted fingerprints stored in his/her database, since \textit{MO} could try to differentiate between encrypted `0's and `1's. 
\end{enumerate}

% % % % % % % % % % % % % % % % % % % % % % % % % % % % % % % % % % % % % % % % % % % % % % % % % % % % % % % % % %

\subsubsection{Collusion attacks}
\label{sec:collusion}
Collusion has been the main research challenge in the realm of fingerprinting. The general weakness of digital fingerprinting occurs when a coalition of buyers compare their uniquely fingerprinted copies of the content to exploit the differences among them trying to remove or alter the fingerprint. The objective of such attack is to evade being traced and, at the same time, possibly frame an innocent buyer. This attack is known as collusion attack and such group of collaborating buyers is called a set of colluders (or a coalition). If a digital fingerprint is not properly designed, a fingerprinting system might fail to detect the traces of any fingerprints under collusion attacks with only a few colluders. To ensure the reliable tracing of true traitors and avoid framing honest buyers, the following collusion attacks are considered in the system:
\begin{enumerate}
	\item \textbf{\textit{Averaging attack.}} In an averaging attack, attackers with a total of $Q$ fingerprinted copies of the same content collude to produce a pirated copy $Y$. The fingerprinted signals are typically averaged with an equal weight for each user:
	\begin{equation*}
	Y(i)= \frac{y_0(i) + y_1(i) +\dots +y_{Q-1}(i)}{Q}.
	\end{equation*}
	\item \textbf{\textit{Minimum attack.}} Under this attack, the colluders create a copy $Y$ whose $i$-th ($i=1,2,\dots,l$) component is the minimum of the $i$-th components of the observed marked copies:
	\begin{equation*}
	Y(i)= \min \left\{y_{0}(i) , y_{1}(i),\dots,y_{Q-1}(i)\right\}.
	\end{equation*}
	\item \textbf{\textit{Maximum attack.}} The colluders create an attacked copy $Y$ by taking the maximum value of the $i$-th components of their individual marked copies:
	\begin{equation*}
	Y(i)= \max\left\{y_{0}(i) , y_{1}(i),\dots,y_{Q-1}(i)\right\}.
	\end{equation*}
\end{enumerate} 
\indent The security of the system against the framing and collusion attacks is discussed in Section \ref{sec:theoretical}.

% % % % % % % % % % % % % % % % % % % % % % % % % % % % % % % % % % % 
\subsection{Architecture of the system}
\label{sec:Model}

\begin{figure}[htbp]
	\centering
	%\graphicspath{G:\Postdoc work\ESWA_david_amna\Fig4.pdf}
	\includegraphics[width=18cm]{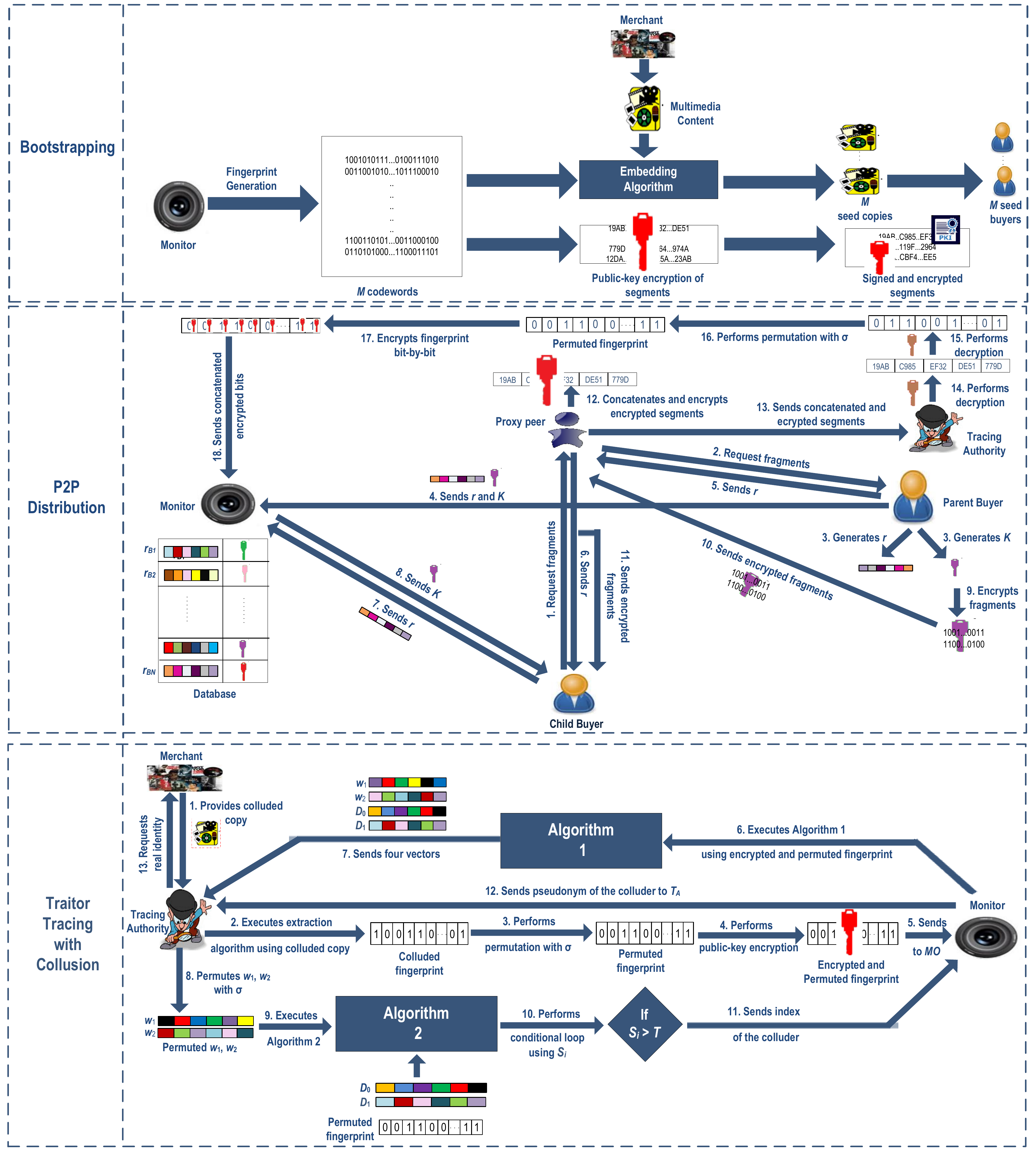}
	\caption{System architecture}
	\label{fig:figh}
\end{figure}

This section describes the architecture of the proposed system. Fig. \ref{fig:figh} shows the structure of the system that contains six main entities: merchant, parent buyer, child buyer, monitor, proxy peers and tracing authority. These entities are involved in three key protocols of the system: bootstrapping, file distribution, and traitor tracing.\\

%%%%%%%%%%%%%%%%%%%%%%%%%%%%%%%%%%%%%%%%%%%%%%%%%%%%%%%%%%
%%%%%%%%%%%%%%%Figure 2%%%%%%%%%%%%%%%%%%%%%%%

In the bootstrapping protocol, \textit{MO} generates Nuida et al.'s codewords for $M$ seed buyers. Then \textit{MO}, executes the embedding algorithm to embed these codewords into the fragments of the multimedia content provided by the merchant. The embedding algorithm produces $M$ seed copies, which are transferred to the seed buyers. Also, \textit{MO} encrypts the segments of the fingerprint codewords with the public key of $T_A$, and then signs these encrypted segments using its secret key.\\

The P2P distribution protocol involves \textit{MO}, the parent buyer, the proxy peers, and the child buyer. The proposed system is configured to select the proxy peers automatically for the content distribution between the parent and the child buyers. When a child buyer requests a content from the merchant, a list of proxy peers is provided to him/her. The child buyer selects the proxy peers from the list to initiate the distribution protocol. After receiving a request from the child buyer, each proxy peer chooses at least two parent buyers, and sends the request to these parent buyers for content transfer. The details of anonymous content transfer between the parent buyers and the child buyer are provided in Protocol \ref{protocol2} in Section \ref{sec:P2Pdistribution}. During the execution of Protocol \ref{protocol2}, each proxy stores the corresponding encrypted segments. After the content is transferred anonymously between the buyers, each proxy peer concatenates the stored segments, encrypts the concatenated segments and sends it to $T_A$. $T_A$ performs the decryption of the encrypted fingerprint with its private key to obtain the encrypted segments, which are re-decrypted by $T_A$ to obtain the cleartext of the fingerprint. $T_A$ sends the permuted and the encrypted fingerprint to \textit{MO}, who stores the received fingerprints in the transaction database. The details of the distribution protocol are provided in Protocol \ref{protocol1}.\\

The traitor-tracing protocol with collusion is performed between the merchant, \textit{MO} and $T_A$. The tracing protocol is initiated by the merchant, who provides the colluded copy to $T_A$. After receiving the colluded copy, $T_A$ extracts the colluded fingerprint from the content using the extraction algorithm. $T_A$ applies permutation and encryption on the colluded fingerprint, and sends it to \textit{MO}. \textit{MO} performs the traitor-tracing algorithm of Nuida et al.'s codes in the encrypted domain and outputs four vectors: $w_1$, $w_2$, $D_0$ and $D_1$ (the details are provided in Algorithm \ref{code2}). \textit{MO} sends these four vectors to $T_A$, who calculates the scores the colluders using these vectors and the permuted fingerprint (details of score calculation are provided in Algorithm \ref{code3}). Once all the scores have been obtained, $T_A$ computes the highest scores $S_i$ that are above a given threshold value ($\mathcal{T}$), and contacts \textit{MO} to reveal the corresponding pseudonym. \textit{MO} sends the pseudonym of the colluder to $T_A$, who then sends the request to the merchant to reveal the real identity of the colluder. \\

\subsection{P2P distribution protocol}
\label{sec:P2Pdistribution}

In the improved system of \citet{Me14}, the fingerprints of the buyers, encrypted with the public keys of both \textit{MO} and $T_A$, are stored by \textit{MO} in its transaction database. The new proposal is to store the fingerprints of the buyers in permuted and encrypted form. $T_A$ permutes and encrypts the fingerprint using a permutation key ($\sigma$) and its own public key ($K_{pA}$), respectively, and then sends the result to \textit{MO}. 

% % % % % % % % % % % % % % % % % % % % % % % % % % % % % % % 
\begin{algorithm}[H]
	\floatname{algorithm}{\textit{Protocol}}
	\caption{\textit{\textbf{(P2P Distribution Protocol)}}}
	\label{protocol1}  
	\begin{algorithmic}[1]
		\item For $i\in \{1,2,\dots,M\}$, \textit{MO} generates the $i$-th seed copy by embedding a \citeauthor{NuFuHaKiWaOgIm07}'s $c$-secure fingerprint \citep{NuFuHaKiWaOgIm07} in each fragment of the content using any transparent, secure and robust watermarking scheme. The details of the fingerprint generation algorithm can be found in \citep{NuFuHaKiWaOgIm07}.
		\item For $i\in \{1,2,\dots,M\}$, \textit{MO} forwards the $i$-th seed copy to the $i$-th seed buyer.
		\item For $i\in \{M+1,M+2,\dots,N\}$, the $i$-th buyer obtains his/her copy of the content by combining the fragments obtained from a set $S_i$ of parent buyers such that $S_i$ $\subseteq$ $\left\{ B_1,B_2,\ldots,B_{i-1}\right\}$ and $|S_i|>1$, where $|\cdot|$ is the cardinality operator. Each transaction is performed via a proxy (or a chain of proxies).
		\State Each fragment of the content contains a fingerprint's segment $g_j$ (with $j=1,\ldots,m$) embedded into it. The fragment is transmitted together with an encrypted version of the segment $E_{K_{pA}}(g_j)$, where $K_{pA}$ is the public key of the tracing authority.
		\item \label{step:protocol2} Each proxy selects a set of $m$ contiguous fragments of the content (Figure \ref{fig:fig2})
		and facilitates the anonymous communication between parents and child for the transmission of those fragments (see Protocol $2$). These $m$ contiguous fragments carry $m$ contiguous segments of the fingerprint embedded into them. After the execution of this step, the proxy stores the corresponding $E_{K_{pA}}(g_j)$.
		\item The proxy concatenates $m$ contiguous encrypted segments, encrypts the concatenation using $K_{pA}$ and sends the result to $T_A$.
		\item $T_A$ performs the decryption on the encrypted fingerprint with its private key $K_{sA}$ to obtain the $m$ contiguous encrypted segments ($E_{K_{pA}}(g_j)$). Once the authority receives all the segments of a fingerprint, $T_A$ further decrypts $E_{K_{pA}}(g_j)$ with $K_{sA}$ to obtain the cleartext of the fingerprint $f_i$. 
		\item $T_A$ uses a secret permutation key $\sigma$ and permutes $f_i$ with $\sigma$ to obtain $\sigma(f_i)$. $T_A$ keeps $\sigma$ since it will be required in the traitor-tracing protocol. Then, $T_A$ encrypts each bit of $\sigma(f_i)$ to produce $E_{K_{pA}}\left(\mathcal{F}\left(\sigma(f_i)^{(j)},r_j\right)\right)$, where $K_{pA}$ is the public key of $T_A$, $\mathcal{F}$ is some ``invertible'' function (possibly public) and $r_i$ is a secret pseudo-random sequence. $\mathcal{F}$ shall be invertible in the sense that there exists some ``inverse'' function $\mathcal{G}$ such that $\mathcal{G}\left(\mathcal{F}(x,r_j),r_j)\right)=x$. Both the function $\mathcal{G}$ and the sequence $r_j$ will be required for decryption.
		\item \label{step:eq1} $T_A$ concatenates the permuted and encrypted fingerprint bits to produce $E_{f_{i}}$ and sends it to \textit{MO}.
		
		\begin{equation}
		\label{eq:1}
		E(\sigma(f_{i}))=E_{K_{pA}}\left(\mathcal{F}\left(\sigma(f_{i})^{(1)},r_1\right)\right) \bigparallel E_{K_{pA}}\left(\mathcal{F}\left(\sigma(f_{i})^{(2)},r_2 \right)\right) \bigparallel \ldots \bigparallel E_{K_{pA}}\left(\mathcal{F}\left(\sigma(f_{i})^{(l)},r_l\right)\right).
		\end{equation}
		Where $\bigparallel$ denotes the concatenation operator.
		\item \label{step:register} The transaction register, created by \textit{MO} for each transaction, stores the following information:
		\begin{tabular}{ll}
			$P_i$: & Pseudonym of the buyer $B_i$\\
			$H(c)$: & Perceptual content hash (used for file indexing) \\
			$E(\sigma(f_{i}))$: & Permuted and encrypted fingerprint of the buyer $B_i$ \\
			$d$: & Transaction date and time (for billing purposes)
		\end{tabular}
	\end{algorithmic}  
\end{algorithm}
% % % % % % % % % % % % % % % % % % % % % % % % % % % % % % % % % % % % % % % % % % % % % % % % % % % % % % % % % %

The new P2P distribution system is detailed in Protocol \ref{protocol1}, from which the following issues can be highlighted:

\begin{enumerate}
	\item  In the first step of the protocol, we can use many different watermarking algorithms, for example \citep{FaMe10} for audio, or \citep{LePrKuMo11} for still images and video. 
	
	\item Eq.~\ref{eq:1}, in Step 9 of the protocol, implies that \textit{MO} cannot decrypt $E(\sigma(f_{i}))$ without the private key ($K_{sA}$) of $T_A$,  and it is also unable to create a new fingerprint by examining the encrypted fingerprints due to an additional layer of security, namely, the permutation of the fingerprint. Also, no single proxy has access to the complete sequence of encrypted segments, since at least two proxies must be chosen by each buyer as described in Section \ref{sec:requirements}. 
	
	\item In Step 8,  the fingerprint is encrypted bitwise by $T_A$, using a a public-key cryptosystem, an ``invertible" function $\mathcal{F}$ and a fixed vector of pseudo-random numbers $r_{j} \in \mathbb{Z}^{*}_{n}$. This vector is generated by $T_A$, with length equal to $l$. The function $\mathcal{F}$ is used combined with the pseudorandom sequence $r_j$ in order to produce different values for binary `1's and `0's at different positions of the fingerprint. 
	
	\item Both $\mathcal{F}$ and the sequence $r_j$ must be the same for the different fingerprints associated to the same content. After decryption, the ``inverse'' function $\mathcal{G}$ and the secret sequence $r_j$ are also required to recover the correct bit, as follows:
	\[
	\mathcal{G}
	\left(D_{K_{sA}}\left(E_{K_{pA}}\left(\mathcal{F}\left(\sigma(f_i)^{(j)},r_j\right)\right) 
	\right),r_j\right)  = \sigma(f_i)^{(j)}.
	\]
	In the sequel, we use $E\left(\sigma(f_i)^{(j)}\right)$ to refer to $E_{K_{pA}}\left(\mathcal{F}\left(\sigma(f_i)^{(j)},r_j\right)\right)$ for the sake of notational simplicity.
	
	\item A possible implementation of this encryption/decryption is the Paillier's cryptosystem \citep{Pail}, which already includes a pseudo-random number for encryption. This choice makes it possible to avoid applying the functions $\mathcal{F}$ and $\mathcal{G}$, since they can be considered to be included in the cryptosystem itself. Because of this, we have taken this approach for implementation, although any other public-key cryptosystem could have been used together with appropriate functions $\mathcal{F}$ and $\mathcal{G}$, and a secret sequence $r_j$.
	
	Paillier is a probabilistic public-key cryptosystem in which a public key is used to encrypt the cleartext, whereas a private key is required to decrypt a ciphertext, and both keys are computationally not easily derived from each other. Paillier is probabilistic in the sense that same cleartext $m_1$ is represented by different ciphertexts. This is due to the fact that in the encryption function of Paillier cryptosystem, a random number $r \in \mathbb{Z}^{*}_{n}$  is chosen so as to produce different ciphertexts for the same cleartext $m_{1}$. The cryptosystem has the property that all the ciphertexts would decrypt to the same cleartext $m_{1}$. This property is very useful when the range of possible cleartext is small, e.g. a single bit data (`0' or `1'). In the absence of such an encryption function, an adversary could statistically determine the proportion of the inputs and gain useful information about the data. 
	With this imlementation, we can simply use $E_{K_{pA}}\left(x,r_j\right)$ and $D_{K_{SA}}\left(y,r_j\right)$ instead of $E_{K_{pA}}\left(\mathcal{F}\left(x,r_j\right)\right)$ and $\mathcal{G}\left(D_{K_{sA}}\left(y\right),r_j\right)$, respectively.
	
	\item However, the encryption of the fingerprints of all the buyers with the same vector $r_{j}$ has a drawback. Since \textit{MO} stores the encrypted fingerprints of all the buyers in its database, it may randomly pick two fingerprints, inspect them and find identical ciphertexts at different positions of these observed fingerprints. Consequently, \textit{MO} could try to infer the possible cleartext value (`0' or `1') of these identical ciphertexts and may attempt to guess some buyer's fingerprint. This drawback is circumvented in our protocol with an additional layer of security, i.e. the permutation of the encrypted fingerprint's bits with $\sigma$. In this way, even if \textit{MO} could infer the values of the cleartext of all the bits of a chose fingerprints of (say) length $l=5,000$ bits, it would still require to compute about $l!=5,000!$ permutations to obtain the valid fingerprint, which would be computationally infeasible.
	
	\item For the anonymous transfer of fragments between buyers in Step 3, Protocol \ref{protocol2} is executed. This is the same four-party anonymous communication protocol used in \citet{Me14} to prevent the proxies from accessing the cleartext of the transferred content between the parents and a child buyer. This protocol protects the symmetric keys shared between the source and the destination buyers using \textit{MO} as a temporary key database. In case a malicious proxy tried to access the database to retrieve the symmetric key, the corresponding record would be erased or blocked, if already retrieved by the legitimate buyer. If the database query of the malicious proxy occurred before that of the legitimate buyer, the latter would find the record either erased or blocked. Hence, the malicious access by the proxy would be detectable. 
\end{enumerate}

% % % % % % % % % % % % % % % % % % % % % % % % % % % % % % % % % % % % % % % % % % % % % % % % % %
\begin{algorithm}[H]
	\floatname{algorithm}{\textit{Protocol}}
	\caption{\textit{\textbf{(Anonymous Content Transfer Between Buyers)}}}
	\label{protocol2}  
	\begin{algorithmic}[1]
		\item The parent buyer chooses a symmetric (session) key $K$.
		\item  The parent chooses a pseudorandom binary sequence $r$ to be used as a handle (primary database key) for $K$. The space for $r$ should be large enough (e.g. $128$ bits) to avoid collisions.
		\item  The parent buyer sends ($r,K$) to \textit{MO}, who stores it in a database.
		\item  The parent buyer sends $r$ to the proxy and the proxy forwards $r$ to the child buyer.
		\item  The child buyer sends the $r$ to \textit{MO}, who replies with the corresponding key $K$.
		\item  \textit{MO} blocks the register ($r,K$) for a given period of time. When the specific time is passed, \textit{MO} removes the register from its database.
		\item  \label{step:encryption} The parent buyer sends the requested fragments, encrypted with $K$, to the proxy.
		\item  The proxy forwards all fragments to the child buyer, who can decrypt them using $K$.
	\end{algorithmic}  
\end{algorithm}
% % % % % % % % % % % % % % % % % % % % % % % % % % % % % % % % % % % % % % % % % % % % % % % % % % % % % % % % % % %

A security analysis of both protocols (\ref{protocol1} and \ref{protocol2}) is provided in Section \ref{sec:formal}.

\subsection{Proposed traitor-tracing protocol}
\label{sec:traitor}

The basic traitor-tracing protocol (without collusion) of the system proposed by \citet{Me14} performs a simple database search to identify an illegal re-distributor. In this protocol, $T_A$ extracts the fingerprint from the pirated copy, encrypts it with the public key of \textit{MO} and its own public key to produce an encrypted fingerprint. This encrypted fingerprint is efficiently searched in the database of \textit{MO}, who reveals the pseudonym of the buyer whose encrypted fingerprint matches the encrypted fingerprint extracted from the pirated copy. 

Similarly, in the new basic traitor-tracing protocol (Protocol \ref{protocol3}) without collusion, $T_A$ first extracts a fingerprint from an illegally re-distributed copy using the same watermarking scheme that was used for embedding the fingerprint in Protocol \ref{protocol1}. Then, $T_A$ permutes the extracted fingerprint $f'$ using $\sigma$, encrypts $\sigma(f')$ with $K_{pA}$ (using also the function $\mathcal{F}$ and the sequence $r_j$), concatenates the permuted and encrypted bits, and sends the result to \textit{MO}. \textit{MO} searches the permuted and encrypted fingerprint in its database and, if found, reveals the  corresponding pseudonym to $T_A$. The real identity of the traitor is disclosed by \textit{Me}, who has an access to the buyers' database. In Protocol \ref{protocol3}, an illegal re-distributor is identified without the involvement of any buyer and, also, without requiring decryption of any segment of the fingerprint. 

\begin{algorithm}[H]
	\floatname{algorithm}{\textit{Protocol}}
	\caption{\textit{\textbf{(Basic traitor tracing without collusion)}}}
	\label{protocol3}  
	\begin{algorithmic}[1]
		\item The fingerprint $f'$ of the illegally re-distributed copy is extracted by $T_A$ using the extraction method and the extraction keys provided by \textit{MO}.
		\item $f'$ is permuted by $T_A$ using $\sigma$ to produce $\sigma(f')$.
		\item Each bit of $\sigma(f')$ is then encrypted with $K_{pA}$ (using also $\mathcal{F}$ and $r_i$), resulting in permuted and encrypted fingerprint $E(\sigma(f'))$ as per Eq. \ref{eq:1}. 
		\item $T_A$ sends $E(\sigma(f'))$ to \textit{MO}.
		\item \textit{MO} searches $E(\sigma(f'))$ in its database. If a matching record is found, the corresponding pseudonym is revealed to $T_A$. Otherwise, an empty set is sent to $T_A$.
		\item If a pseudonym is obtained, $T_A$ sends a request to \textit{Me} to provide the real identity of the traced buyer. Then, \textit{Me} checks his/her database of buyers and retrieves the real identity and \textit{AGR} corresponding to the pseudonym received from $T_A$. If no pseudonym is obtained, the traitor-tracing protocol with collusion (Protocol \ref{protocol4}) must be run.
	\end{algorithmic}  
\end{algorithm}

In case a traitor is not identified in Protocol \ref{protocol3}, the traitor-tracing protocol with collusion (Protocol \ref{protocol4}) is executed to identify a colluder or a set of colluders. The traitor-tracing protocol of \citep{Me14}, in case of collusion of several buyers, requires a decrypted fingerprint's hash to trace a colluder. In this regard, \textit{MO} requires the cooperation of the involved proxies to construct a valid codeword for each hash when the content is  distributed to each buyer. The hash $h(f')$ of a colluder can be reconstructed from the fingerprint $f'$ extracted from the colluded copy. This hash is then encrypted using the public key of \textit{MO} to produce $E_{K_{p_{MO}}}(h_{f'})$. Then, $E_{K_{p_{MO}}}(h_{f'})$ is searched in the database of \textit{MO}, instead of the encrypted fingerprint. As described in \citep{Me14}, with a large enough hash space, hash collisions would be almost negligible and a colluder would still be identified in the vast majority of the cases.

% % % % % % % % % % % % % % % % % % % % % % % % % % % % % % % % % % % 
\begin{algorithm}[H]
	\floatname{algorithm}{\textit{Protocol}}
	\setcounter{algorithm}{3}
	\caption{\textit{\textbf{(Traitor tracing with collusion)}}}
	\label{protocol4}  
	\begin{algorithmic}[1]
		\item  $T_A$ extracts the colluded fingerprint $f'$ from the content, permutes it using the permutation key $\sigma$, encrypts it bitwise, obtaining $E(f')$, and sends it \textit{MO}.
		\item \textit{MO} computes four vectors $\omega_{1}$, $\omega_{2}$, $D_{0}$, $D_{1}$ using Algorithm \ref{code2} for all buyers $B_i$, and sends them to $T_A$.
		%\item \textit{MO} sends these vectors to $T_A$.%, who further carries out the traitor-tracing protocol by computing scores of the buyers.
		\item $T_A$ performs the permutation of the received $\omega_{1}$, $\omega_{2}$, using $\sigma$, in Algorithm \ref{code3}.
		\item The score $S_i$ of the each buyer $B_i$ is computed by $T_A$, using $D_{0}$, $D_{1}$, and the permuted $\omega_{1}$, $\omega_{2}$ and $f'$, in Algorithm \ref{code3}.
		\item $T_A$ obtains the indexes of the colluders (at least one) with the highest scores (greater than the global threshold $\mathcal{T}$) and sends them to \textit{MO}. 
		\item \textit{MO} sends the pseudonyms of the colluders to $T_A$ (corresponding to the indexes of the previous step). The real identities of the buyers is not known by $T_A$, only the pseudonyms are revealed. 
		\item In order to obtain the real identity of the colluders, $T_A$ contacts \textit{Me}, since only he/she has the access to buyers' database, who associates the pseudonyms with the corresponding real identities. \textit{Me} also provides a proof (\textit{AGR}) of the association between the pseudonyms and the identities.
	\end{algorithmic}
\end{algorithm}
% % % % % % % % % % % % % % % % % % % % % % % % % % % % % % % % % % % 

In the new traitor-tracing system with collusion, $T_A$ extracts the fingerprint $f'$ from the colluded copy, permutes $f'$ using $\sigma$, and encrypts the result with $K_{pA}$ (using also $\mathcal{F}$ and $r_j$) to produce $E(\sigma(f'))$ as per Eq. \ref{eq:1}. Then, $T_A$ sends $E(\sigma(f'))$ to \textit{MO}, who performs the first part of the traitor-tracing algorithm (Algorithm \ref{code2}) of \citeauthor{NuFuHaKiWaOgIm07}'s fingerprinting codes. 

The standard  traitor-tracing algorithm published by \citet{NuFuHaKiWaOgIm07} is carried out in the cleartext domain by comparing the traced and the tested fingerprints, and outputs the set of colluders with the highest scores. In the proposed system, due to the requirement of buyer frameproofness (see Section \ref{sec:requirements}), \textit{MO} stores the permuted and encrypted fingerprints in his/her transaction database. Thus, in order to perform \citeauthor{NuFuHaKiWaOgIm07}'s traitor-tracing algorithm in cleartext, \textit{MO} would need the private key of $T_A$ (and also the sequence $r_j$ and the function $\mathcal{G}$) together with $\sigma$. Since $T_A$ is the trusted party of the system, only it can perform operations on a cleartext fingerprint. Therefore, in order to protect the frameproofness of innocent buyers, it is required that \textit{MO} performs the traitor-tracing algorithm of \citeauthor{NuFuHaKiWaOgIm07}'s codes in the encrypted domain, in such a way that the scores of the colluders obtained in the encrypted domain match those that would be calculated in the cleartext domain.

% % % % % % % % % % % % % % % % % % % % % % % % % % % % % % % % % % % 
\begin{algorithm}[ht]
	%\newfloat{algorithm}{tbp}{}
	\setcounter{algorithm}{0}
	\caption{Traitor-tracing algorithm performed by \textit{MO}}
	\label{code2}
	\begin{algorithmic}[30]
		\Function{\citeauthor{NuFuHaKiWaOgIm07}'s Traitor-tracing Algorithm}{}
		\item {Input parameters: $E(\sigma(f'))$, $p$, $E(\sigma(f_{i}))$}
		\item {Output parameters: $\omega_{1}$, $\omega_{2}$, $D_{0}$, $D_{1}$}
		\item {Use function $\phi(x) \doteq \displaystyle \sqrt{(1-x)/x}$, for $x\in[0,1]$.}
		\item \textbf{begin}
		\For {$k\in\{1,2,\dots,l\}$} 
		\State $\omega^{(k)}_1 \gets \phi(1-p^{(k)})$
		\Comment{Calculate two vectors of the scores $\omega^{(k)}_{1}$ and $\omega^{(k)}_{2}$ using $\phi(p)$}\\
		\State $\omega^{(k)}_2 \gets \phi(p^{(k)})$
		\If {$E\left(\sigma(f')^{(k)}\right)=E\left(\sigma(f_{i})^{(k)}\right)$}
		\State $D^{(k)}_{0} \gets  +1$  
		\State $D^{(k)}_{1} \gets  +2$
		\Else
		\State $D^{(k)}_{0} \gets -2$   
		\State $D^{(k)}_{1} \gets -1$
		\EndIf
		\EndFor
		\State \Return $\omega_{1}$, $\omega_{2}$, $D_{0}$, $D_{1}$
		\EndFunction
	\end{algorithmic}
\end{algorithm}
% % % % % % % % % % % % % % % % % % % % % % % % % % % % % % % 

% % % % % % % % % % % % % % % % % % % % % % % % % % % % % % % % % % % 
\begin{algorithm}[ht]
	%\newfloat{algorithm}{tbp}{}
	\setcounter{algorithm}{1}
	\caption{Traitor-tracing algorithm performed by $T_A$}
	\label{code3}
	\begin{algorithmic}[30]
		\Function{\citeauthor{NuFuHaKiWaOgIm07}'s Traitor-tracing Algorithm}{}
		\item {Input parameters: $\omega_{1}$, $\omega_{2}$, $D_{0}$, $D_{1}$, $f'$, $\sigma$}
		\item {Output parameter: $S_{i}$}
		\item {Use function $\mathrm{sign}(x) \doteq \left\{ 
			\begin{array}{cl}
			1,  &  \text{if } x \geq 0, \\
			-1, &  \text{otherwise.}
			\end{array}\right.$}
		\item \textbf{begin}
		%\State $f'^{(k)} \gets \sigma(f'^{(k)})$
		\State $W_{1} \gets \sigma(\omega_{1})$
		\State $W_{2} \gets \sigma(\omega_{2})$
		\For {$k\in\{1,2,\dots,l\}$} 
		\State $j\gets{\sigma(f')^{(k)}}$ 
		%\State $\mathcal{W}^{(k)} \gets  \left\{W^{(k)}_{1},W^{(k)}_{2}\right\}$\\
		
		\State $S^{(k)}_{i} \gets{\mathrm{sign}\left(D^{(k)}_j\right) \cdot W^{(k)}_{\left|D^{(k)}_j\right| } }$
		% _{\}} \\
		\EndFor
		\State $S_{i} \gets \displaystyle{ \frac{1}{l} \sum_{k=1}^{l}S^{(k)}_i }$
		\State \Return $S_{i}$ 
		\EndFunction
	\end{algorithmic}
\end{algorithm}
% % % % % % % % % % % % % % % % % % % % % % % % % % % % % % % 

In the standard \citeauthor{NuFuHaKiWaOgIm07}'s traitor-tracing algorithm, the scores are calculated on the basis of bit differences or similarities between the traced fingerprint $f'$ and the tested fingerprint $f_i$. The scores obtained through the modified \citeauthor{NuFuHaKiWaOgIm07}'s traitor-tracing protocol (i.e. in the encrypted domain) are identical to those obtained using standard \citeauthor{NuFuHaKiWaOgIm07}'s traitor-tracing algorithm. The detailed description of the modified \citeauthor{NuFuHaKiWaOgIm07}'s traitor-tracing protocol is provided in Protocol \ref{protocol4}. It can be noticed that part of the computations (Algorithm \ref{code2}) are carried out by \textit{MO} and the rest are performed by the tracing authority (Algorithm \ref{code3}). The reason for this is that only the authority knows the permutation key $\sigma$ used to shuffle the bits of fingerprints, in such a way that the cleartext of the fingerprints are never available for the monitor, yielding buyer frameproofness.

Since \textit{MO} needs to perform the traitor-tracing algorithm in the encrypted domain, the scores will be calculated using $E(\sigma(f'))$, $E(\sigma(f_i))$ and $p$ as inputs, where $p$ is the secret vector used by \textit{MO} in the generation of the segment codewords. Since traitor-tracing is performed using the permuted fingerprints, $p$ should also be permuted in order to compute the scores of the buyers. However,  \textit{MO} cannot carry out this permutation, because $\sigma$ is known only by $T_A$. A possibility to overcome this difficulty would be to send $p$ to $T_A$, who would perform the permutation on $p$ and send the result back to \textit{MO}. However, this solution would not be viable, since \textit{MO} could observe the differences between $p$ and $\sigma(p)$, infer $\sigma$, and try to use the permutation key to disclose a valid fingerprint. Another possibility would be to send $p$ and $E(\sigma(f_{i}))$ from \textit{MO} to $T_A$, who would perform permutation on $p$ and then calculate the scores of the buyers. However, this exchange of encrypted fingerprints (the whole database) would not be efficient, since it would require enormous communication costs (e.g. if milions of users have purchased the traced content). Also, as discussed in Section \ref{sec:entities}, traitor-tracing is considered more effective and reliable when this functionality is performed by an entity different from the party responsible for the generation of the fingerprints. 

In the proposed traitor-tracing algorithm with collusion (Algorithm \ref{code2}), \textit{MO} calculates two vectors, $\omega_1$ and $\omega_2$, using vector $p$, instead of sending $p$ to $T_A$. Obviously, $T_A$ could compute $p$ from $\omega_1$ and $\omega_2$, but $T_A$ is a trusted entity (Section \ref{sec:assumptions}) and it is not expected to behave maliciously. Then, \textit{MO} compares the encrypted bits $E(\sigma(f_{i}))$ and $E(\sigma(f'))$  and outputs two vectors ($D_{0}$ and $D_{1}$), which contain pairs of either positive or negative indexes of $\omega_1$ and $\omega_2$. Later on, $D_0$ will be used when the corresponding cleartext is `0', whereas $D_1$ is used for a cleartext equal to `1'. 

In the original \citeauthor{NuFuHaKiWaOgIm07}'s traitor-tracing algorithm, in case of identical bits in $f_{i}^{(k)}$ and $f'^{(k)}$ (either both `1's or both `0's), a positive score equal to $\omega^{(k)}_{1}$ or $\omega^{(k)}_{2}$ is assigned to that bit position. On the other hand, a differing bit between $f_{i}^{(k)}$ and $f'^{(k)}$ results in a negative score, either $\omega^{(k)}_{1}$ or $\omega^{(k)}_{2}$. The same approach is adopted in Algorithm \ref{code2}, where in case of an equality result between $E(\sigma(f')^{(k)})$ and $E(\sigma(f_{i})^{(k)})$,  \textit{MO} assigns  a value with a positive sign (indicated by `+') to the ``subscript vectors'' $D^{(k)}_{0}$ and $D^{(k)}_{1}$. Similarly, in case of inequality, a negative value (indicated by `$-$') is assigned to the ``subscript vectors''. In the final step of Algorithm \ref{code2}, \textit{MO} returns the output ($\omega_{1}$, $\omega_{2}$, $D_{0}$ and $D_{1}$) that will be sent to $T_A$, who further carries out the remaining steps of the original \citeauthor{NuFuHaKiWaOgIm07}'s traitor-tracing algorithm (as described in Algorithm \ref{code3}).

In Algorithm \ref{code3}, $T_A$ computes a score of the buyer with $\omega_{1}$, $\omega_{2}$, $D_{0}$, $D_{1}$ and $f'^{(k)}$ and $\sigma$ as its inputs. Here, the cleartext bits $f'^{(k)}$ are required by $T_A$ in order to determine the correct score for each bit position, which will make it possible to select the appropriate value $\omega^{(k)}_{1}$ or $\omega^{(k)}_{2}$. The absolute value of $D^{(k)}_{0}$ or $D^{(k)}_{1}$ is used as a subscript for the selection of the appropriate value, whereas the sign of $D^{(k)}_{0}$ or $D^{(k)}_{1}$ is used to determine the sign of the score of that particular bit. It is worth pointing out that $T_A$ first performs the permutation $\sigma$ on $f'$, $\omega_{1}$ and $\omega_{2}$ to yield $\sigma(f')$, $W_{1}$ and $W_{2}$, respectively.

%It then constitutes a vector $\mathcal{W}^{(k)}$, which contains pairwise values for bit differences or similarity between an encrypted traced fingerprint and an encrypted tested fingerprint. $\mathcal{W}$ can be expressed in the following form:

%\begin{equation*}
%\mathcal{W}=\left\{(W_{1,1},W_{2,1}),(W_{1,2},W_{2,2}),\ldots,(W_{1,l},W_{2,l})\right\}
%\end{equation*}

In short, $T_A$ computes the scores of a buyer bit-by-bit by using the bits of $\sigma(f')^{(k)}$ in the following way: If $\sigma(f')^{(k)}$ is equal to `0', $T_A$ selects $D^{(k)}_{0}$, which points towards the value contained in either $W_{1}^{(k)}$ (positive case) or $W_2^{(k)}$ (negative case), depending on the result of the comparison carried out by \textit{MO} in Algorithm \ref{code3}, along with its sign. 
Alternatively, if $\sigma(f')^{(k)}=\text{`1'}$, $T_A$ selects $D_{1}^{(k)}$, which refers to a positive value stored in $W_{2}$ or a negative value stored in $W_{1}$, to compute the score for that bit. When all bitwise scores of a buyer have been computed, $T_A$ calculates the total score of that buyer by taking the average of the all bitwise scores. 

Finally, once all the scores for all buyers have been obtained, $T_A$ computes the highest scores (which shall be above a given threshold $\mathcal{T}$) and contacts \textit{MO} to reveal the corresponding pseudonyms. Simulated results and a theoretical analysis of the traitor-tracing protocol with collusion are provided in Section \ref{sec:simulated}.

\section{Features of the proposed system and comparative analysis}
\label{sec:Features}
In this section, we discuss how the drawbacks of \citep{Me14} are circumvented by the new protocols and algorithms. Besides, we compare the features and shortcomings of existing anonymous fingerprinting schemes for P2P networks with the proposed system.

\subsection{Features of the proposed system}
\label{sec:FeaturesPS}
As discussed in Section \ref{sec:drawbacksMe14}, the system proposed by \citet{Me14} exhibits two major drawbacks: 1) the fingerprinting scheme requires a two-layer anti-collusion code, which results in a longer codeword, and 2) the construction of a valid hash-level anti-collusion code requires the participation of the proxies, supervised by \textit{MO}, with a non-specified protocol to guarantee that the hash-level codeword is valid. Both drawbacks are caused by the requirement of the two-layer fingerprint encoding in order to obtain collusion resistance. 

In the proposed work, this first drawback is overcome by using the state-of-the-art \citeauthor{NuFuHaKiWaOgIm07}'s codes in a segment-wise form. \citeauthor{NuFuHaKiWaOgIm07}'s codes provide outstandingly short code lengths. For example, 10 different codewords providing collusion resistance against $c\leq c_0=4$ colluders can be obtained with codewords of length $l_0=788$ bits. If $l_0$ is the length of a codeword with collusion resistance against $c \leq c_0$ colluders, the total fingerprint will have length $l=l_0n_s$, where $n_s$ is the number of segments of the fingerprint, as shown in Figure \ref{fig:fig2}.  If we build a fingerprint with 74 segments to provide collusion resistance against $c\leq4$ colluders, with 10 different codewords (i.e. the number of seed buyers should be no greater $M\leq 10$), we require fingerprints with $l=788\cdot74=58,312$ bits, much shorter than those required by the systems in \citep{MeDo13,Me14}.  Note that 74 segments is more than enough to guarantee the numerical explosion of the fingerprint space. If 10 different values (codewords) are available for each segment (as required if we have $M=10$ seed buyers), the number of possible fingerprints becomes $10^{74}$, many orders of magnitude larger than the population of the Earth. 

On the other hand, in \citep{Me14}, the length of the fingerprint is obtained by multiplying the length of the segment-level code by the length of the hash-level code, which may result in too long fingerprints for real applications. E.g. if the length of the segment-level codewords is, again, 788 bits (providing resistance against $c\leq4$ colluders and 10 different codewords), and that of the hash-level codewords is $1,577$ bits for resistance against $c\leq4$ colluders and $100,000$ different codewords (which would be the maximum number of buyers for that content), the whole fingerprint would be $788\cdot 1,577 = 1,242,676$ bits long, which could be unsuitable for many applications (e.g. relatively short music files). Note that $58,312$ bits are enough for $10^{74}$ buyers with the proposed system, whereas $1,242,676$ bits only suffice for $100,000$ buyers with the system of \citep{Me14}, in both cases providing resistance against $c\leq4$ colluders.

Similarly, the second drawback is overcome by the proposed system since any recombination of segments produces a valid fingerprint. Thus, we no longer require the construction of a hash-level anti-collusion codeword among segments.  In \citet{Me14}, a valid hash-level codeword must be constructed with the help of the proxies and the supervision of \textit{MO}. In order to obtain a valid codeword, the proxies need to contact more parents, due to the fact that the hash-level collusion-resistant code may require a specific hash bit (`0' or `1') for a particular position. If no parent is available with that particular hash bit, the proxy must wait for an undetermined period until a valid segment becomes available from some parent.

The construction of the fingerprints is, thus, much simpler in the proposed system, since any recombination of segments yields a valid fingerprint. Hence, the new distribution protocol does not require the verification of the fingerprint. 
In addition, in the proposed distribution protocol, \textit{MO} stores the fingerprint of each buyer permuted and encrypted by $T_A$, which is a trusted party, using $\sigma$ and its public key ($K_{pA}$). A signature of both the segment and the corresponding content's fragment can be transmitted alongside the segment to prevent proxies from cheating. This signature approach is also used in \citep{Me14} and \citep{MeDo13} to prevent the proxies from reporting fake segments in the distribution protocol.

Finally, a new protocol for traitor tracing (Protocol \ref{protocol4}) is proposed, in which guilty buyers with the highest scores (above some threshold) are successfully traced and identified. The standard \citeauthor{NuFuHaKiWaOgIm07}'s traitor-tracing system has been modified to work in the encrypted domain (yielding the same result as the cleartext counterpart). Hence, the traitor-tracing protocol is executed without requiring \textit{MO} or $T_A$ to decrypt any single segment of the buyers' fingerprints and, also, without requiring the cooperation of any single buyer of the system.

% to overcome the drawback of the referred system \citep{Me14}. is based on \citeauthor{NuFuHaKiWaOgIm07}'s traitor-tracing algorithm, 

%Each segment of the fingerprint is embedded with \citeauthor{NuFuHaKiWaOgIm07}'s codes and provides resistance against $c=4$ colluders. 

% % % % % % % % % % % % % % % % % % % % % % % % % % % % % % % % % % % 
\subsection{Comparative analysis}
\label{sec:comparative}
In this section, we provide a comparative analysis of the properties of the suggested scheme against  existing recent anonymous fingerprinting schemes for P2P systems \citep{DoMe13,MeDo13, Me14,qmr15,qmr16}. The comparison, summarized in Table \ref{tab:comparative}, considers the following desirable properties for large-scale P2P anonymous fingerprinting schemes:
\begin{enumerate}
	\item \textbf{Embedding required for only a few fingerprints:} Whether the system works by embedding only a small number of fingerprints. The smaller the number of embeddings, the most efficient the distribution becomes either for the merchant or for the peer buyers (or both).
	\item \textbf{Guaranteed traitor tracing:} Whether the system guarantees traitor tracing when the number of colluders $c$ does not exceed the collusion resistance capacity of the fingerprinting code ($c\leq c_0$).
	\item \textbf{Single-level anti-collusion code:} Whether the system works with a single-level anti-collusion code or requires several levels of coding.
	\item \textbf{Fully P2P distribution system:} Whether the distribution of all the files occurs in a fully P2P manner. 
	\item \textbf{Avoids public-key encryption of contents:} Whether the system avoids public-key encryption of the contents.
	\item \textbf{Efficient protocols:} Whether the protocols are efficient (avoiding costly mechanisms, such as zero-knowledge proofs or secure multi-party computations).
\end{enumerate}

\begin{table}[ht]
	\centering
	\begin{tabular}{||m{2.2cm}||m{1.7cm}|m{1.7cm}|m{1.7cm}|m{1.7cm}|m{1.7cm}|m{1.7cm}||}
		\hhline{~|t:======:t|} 
		\multicolumn{1}{c||}{} & Embedding required for only a few fingerprints & Guaranteed traitor tracing & Single-level anti-collusion code & Fully P2P distribution system & Avoids public-key encryption of contents & Efficient protocols \\ 
		\hhline{|t:=::======:|} 
		\cite{DoMe13} & No & Yes & Yes & No & Yes & No \\ \hhline{||-||------||} 
		\cite{MeDo13} & Yes & No & No & Yes & Yes & Partially \\ \hhline{||-||------||}   
		\cite{Me14} & Yes & Yes & No & Yes & Yes & Partially \\ \hhline{||-||------||}   
		\cite{qmr15} & No & Yes & Yes & No & Partially & Yes, partially centralized \\ \hhline{||-||------||}   
		\cite{qmr16} & Yes & Yes & Yes & No & Yes & Yes, partially centralized \\ \hhline{||-||------||}   
		\textbf{Proposed} & \textbf{Yes} & \textbf{Yes} & \textbf{Yes} & \textbf{Yes} & \textbf{Yes} & \textbf{Yes} \\ \hhline{|b:=:b:======:b|} 
	\end{tabular}
	\caption{Comparative analysis of P2P-based anonymous fingerprinting schemes}
	\label{tab:comparative}
\end{table}

The comparative analysis of Table \ref{tab:comparative} points out the following issues:
\begin{itemize}
	\item The system of \citep{DoMe13} requires a specific embedding per each buyer, although this embedding is carried out by peer buyers except for the first generation, for which the merchant is the party who embeds the fingerprint. The main drawbacks of this system are: 1) it does not provide a fully P2P distribution scheme, since the transfers are carried in tree form (not a graph), and 2) it requires a non-specified complex and costly secure multi-party computation protocol for each file transfer to guarantee frameproofness.
	\item The scheme proposed in \citep{MeDo13} entails two major drawbacks: 1) the co-operation of innocent buyers in traitor tracing may lead to situations in which an illegal re-distributor cannot be traced by the authority, and 2) it requires a double layer anti-collusion code (one for the fingerprint segments and another one for the fingerprint hash). To create the hash level codeword, the proxies and the monitor will have to deploy a complex protocol in such a way that the codeword is not revealed to any single party. This protocol is unlikely to be efficient (it will require multi-party secure computation or a similar approach).
	\item The system of \citep{Me14} solves most of the drawbacks of the previous scheme, but still requires the two-layer encoding of the fingerprint, which entails the problems discussed above.
	\item The proposal of \citep{qmr15} requires the merchant to participate in the distribution protocol and embed the fingerprint in a short-sized base file (for each single buyer), whereas a much larger supplementary file --useless without the former-- is distributed through the P2P network. The distribution of the base file requires embedding the fingerprint in the encrypted domain and, hence, the contents are partially encrypted (using homomorphic encryption). As a result, the merchant is involved in computations and transfers for each single purchase, and hence the distribution is not purely P2P.
	\item Recently, the system proposed in \citep{qmr16} uses some of the ideas of \cite{qmr15}, but replaces homomorphic encryption by fragmentation, symmetric encryption, permutation and distribution of the fragments through a collection of proxy peers. In addition, the embedding is pre-computed by embedding a string of all 1's and one of all 0`s in the original content. The appropriate bit for each coefficient is selected by the proxies from the pre-computed values. Although the protocol for the creation and distribution of the base file is much lighter than the one of \cite{qmr15}, it still requires the participation of the merchant for each file transfer. Hence, the distribution is not fully P2P. The P2P network is used only for the distribution of the large-sized supplementary file. Therefore, as scalability is concerned, the computational and communication costs for the merchant will grow linearly with the number of buyers.   
	\item Finally, the system proposed in this paper provides the most scalable and efficient solution. The merchant only needs to embed the fingerprint for the seed buyers, whose number ($M$) will be typically small (e.g. around 10). Thereafter, all the transfers are carried out in a purely P2P fashion, and the recombination of the segments of the fingerprints is enough to produce unique identifiers of buyers due to numerical explosion. Public-key encryption is required only for short strings and not for the multimedia contents. The introduction of \citeauthor{NuFuHaKiWaOgIm07}'s codes makes it possible to work with a single-level collusion-resistant code, resulting in a simple P2P distribution protocol. The standard traitor-tracing algorithm has been replaced by a protocol between the monitor and the authority, in which permuted and encrypted fingerprints are used, making it possible to distinguish between colluders and innocent buyers. The operation of such protocol is analyzed in the following section, both theoretically and by means of simulations.
\end{itemize}

Among the analyzed systems, the proposed work is the one that provides the best properties. After bootstrapping, all transfers occur within the P2P network without any participation of the merchant. This provides excellent efficiency and scalability, with reduced costs for the merchant, which may give rise to
cheaper prices for the buyers. Consequently, in this paper, we successfully overcome the results of prior art  \citep{MeDo13,Me14,qmr15,qmr16} The new proposal is much more practical and easier to implement. Hence, if digital media producers decided to adopt this paradigm, they would not be afraid of the illegal usage and re-distribution of their products, making it possible to fight against the reluctance to use P2P networks due to piracy. Therefore, we provide a much more efficient, scalable and cheap solution for content providers.

% % % % % % % % % % % % % % % % % % % % % % % % % % % % % % % % % % % 
% % % % % % % % % % % % % % % % % % % % % % % % % % % % % % % % % % % 
\subsection{Discussion}
\label{sec:Discuss}

This section discusses the strengths, weaknesses, and limitations of the proposed fingerprinting system.\\

\par{\textbf{\textit{A. Strengths of the system:}} The notable features of the proposed system that make it novel and effective are the following:}\\

\begin{itemize}
	\item Distribution produces automatic fingerprinting by recombination with no additional embedding. After bootstrapping, we only need to recombine the existing fingerprints in each transaction.
	\item The computational and communication costs for the merchant are zero per each transaction, which is quite novel in fingerprinting scenarios. Once the system is started, the merchant does not need to participate in any transfer.
	\item The proposed scheme prevents the double layer encoding of the fingerprint required by the previous recom-bination-based schemes \citep{MeDo13,MeDo13a,Me14}. This two-layer encoding was very problematic, since it produced long fingerprints and also required supervision by the monitor in the distribution system (not all recombined fingerprints were valid).
\end{itemize}

\par{\textbf{\textit{B. Disadvantages of the system:}} This part of the section discusses the weaknesses of the system, and also the countermeasures to mitigate these disadvantages.}

\begin{itemize}
	\item The first weakness of the system is related to the anonymous communication protocol between peer buyers that uses onion routing-like protocol, which causes a cryptographic overhead. This overhead is due to encryptions and decryptions, and increase in the routing path to provide anonymity between two communicating parties. Any system that provides anonymous communication to prevent direct connection between a sender and a receiver has to bear the costs arising from encryption/decryption operations and increased routing traffic. To mitigate this weakness, we used symmetric cryptography instead of asymmetric to reduce the cryptographic overhead. 
	\item The second disadvantage is related to the size of a fingerprint, which is shorter than in previous recombined approaches, but still can have a size of several kilobytes, resulting in a database of fingerprints of several megabytes (MBs) or even gigabytes (GBs) for each content. However, with the advancement in the storage technology and reduced costs, we envision high capacity storage devices in the deployment of our system that could easily store fingerprints databases of the order of GBs.
	\item The third weakness of the system corresponds to tracing colluder(s) in an encrypted domain in traitor-tracing protocol (Protocol \ref{protocol4}), which causes computational overhead for the tracing authority. To subdue this issue, a cloud-based service could be used, that would be solely responsible for executing the traitor-tracing protocol. Since the traitor-tracing protocol with collusion (Protocol \ref{protocol4}) is executed in an encrypted domain, the use of cloud would not raise any security or privacy concerns.
\end{itemize}

\par{\textbf{\textit{C. Limitations of the system:}} In the following, we discuss the constraints of the proposed system with respect to the anonymous fingerprinting scheme:}
\begin{itemize}
	\item The recombination scheme works as long as there are at least two parents available per each transaction. Other-wise, the fingerprint obtained by a child would be identical to that of his/her only parent, what is unsuitable for traitor tracing.
	\item Randomness is required for a correct recombination of segments. The proxies should not simple take the first half fragments from a parent and the second half fragments from a second one. This would produce identical fingerprints during distribution and, hence, the traitor-tracing protocol would result in incorrect results. For example, if a proxy chooses two parents for $10$ fragments, he/she may select fragments $1$, $4$, $6$, $8$ and $9$ from one parent, and $2$, $3$, $5$, $7$ and $10$ from the other one, instead of choosing $1-5$ and $6-10$. This must be carefully enforced in the distribution protocol. Malicious proxies deviating from the distribution protocol should be detected to prevent repeated fingerprints.
	\item Also, the proxies should not deviate from the protocols in other key steps, such as those required to store the encrypted fingerprints in the transaction monitor's database. Although the protocols introduce controls to detect malicious proxies, there must be a constant supervision of the proxies to detect them quickly and prevent a flawed fingerprint database that could undermine the tracing system. 
	\item Although $n_s$ segments can yield, theoretically, $M^{ns}$ different fingerprints, this is not really true, since two fingerprints differing only in a few segments would be considered identical by the traitor-tracing protocol. This is also related to the second weakness of this list. The larger the number of segments ($n_s$), and the larger the number of seed buyers ($M$), the probability of having two equivalent fingerprints becomes vanishingly small. With the numbers selected for the experiments ($n_s = 74$ and $M = 10$) and a good random selection of the fragments, two equivalent fingerprints would be almost impossible even for billions of buyers (since $10^{74}$ is a huge number compared to $10^9$).
	\item Finally, the fragments should be transferred completely from parent to child (through the proxies). The fragments can be further re-fragmented by the network protocols, but each fragment should be reconstructed completely at the child’s end. 
\end{itemize}

\section{Theoretical analysis and simulation results}
\label{sec:theoretical}
This section analyzes how the design goals (described in Section \ref{sec:design}) of the proposed system are achieved. Firstly, formal and informal analyses concerning the correctness of the protocols of the proposed system, in terms of security and privacy, are presented. Secondly, the novel traitor-tracing protocol presented in Section \ref{sec:traitor} is examined both theoretically (in order to derive the appropriate global threshold $\mathcal{T}$) and by means of simulations. The effectiveness of the traitor-tracing mechanism against linear (averaging) and non-linear (maximum and minimum) collusion attacks is presented in Section \ref{sec:collusion} using simulated experiments.

\subsection{Security and privacy analysis}
\label{sec:secanalysis}
In this section, formal and informal discussions are provided for the security and privacy of the system according to the design requirements and the attack model presented in Sections \ref{sec:design} and \ref{sec:model}.

\subsubsection{Formal analysis of the protocols}
\label{sec:formal}
Formal theorems and their corresponding proofs are provided in this section to analyze the security of the proposed protocols: P2P distribution (Protocol \ref{protocol1}), anonymous communication (Protocol \ref{protocol2}) and traitor tracing (Protocols \ref{protocol3} and \ref{protocol4}).

\begin{thm}
	Malicious parties cannot frame an innocent buyer in Protocol \ref{protocol1}.
\end{thm}
\begin{pf}
	There would be two ways of framing an innocent buyer $B_i$ in Protocol \ref{protocol1}:
	\begin{enumerate}
		\item If another party gains access to the cleartext content of that buyer and re-distributes it.
		\item If the fingerprint $f_j$, corresponding to a buyer $B_j$ with $j\neq i$, is linked to the pseudonym of buyer $B_i$ instead of $B_j$ in the transaction database. If $B_j$ re-distributed his/her content later on, the pseudonym of $B_i$ would be retrieved instead of that of $B_j$.
	\end{enumerate}
	Both attacks would require the participation of the transaction monitor \textit{MO}, which is a trusted party for the management of the session keys and the transaction database. The infeasibility of these two alternatives is discussed below. \begin{enumerate}
		\item The transmission of the fragments of the content from parents to a child occurs in Step \ref{step:protocol2} of Protocol \ref{protocol1}. This transmission is carried out as detailed in Protocol \ref{protocol2}. In this protocol, the fragments are transmitted from parents to child encrypted with different session keys ($K$). These keys are managed and protected by \textit{MO} as described in the steps of  Protocol \ref{protocol2}. The monitor \textit{MO} is a trusted party in Protocol \ref{protocol2} for the management of the session keys (see the security assumptions in Section \ref{sec:assumptions}). Hence, this kind of attack is not possible due to the trustiness of \textit{MO} for the management of the session keys.
		\item The transaction register, created in Step \ref{step:register} of Protocol \ref{protocol2}, contains the fingerprint of each buyer previously encrypted and permuted by the authority ($E(\sigma(f_i))$, for all $i$). \textit{MO} cannot link the fingerprint $f_j$ of the buyer $B_j$ to a different buyer $B_i$, for some $j\neq i$, since, as detailed in the security assumptions, the monitor is trusted for the management of the transaction database.
	\end{enumerate}
	Both possibilities are thus discarded due to the security assumptions related to \textit{MO}. Any collusion of proxies and the monitor can also be discarded by the same security assumption (trustiness of \textit{MO}).
	\qed
\end{pf}
% In principle, \textit{MO} could collude with all the proxies of a transfer in Protocol \ref{protocol1} to obtain some buyer's fingerprinted copy of the content (which could be re-distributed illegally afterwards). The fingerprinted copy could also be used to extract the complete fingerprint from it, since \textit{MO} has access to the extraction algorithm and keys.

% Since \textit{MO} manages the session keys, it could decrypt the content fragments if they are forwarded to it by malicious proxies. By joining all the fragments, the complete fingerprinted copy would be obtained by \textit{MO}. Such coalition, however, can be discarded since \textit{MO} is trusted for the management of the session keys (see the security assumptions in Section \ref{sec:assumptions}).\qed

\begin{rmk}
	Even if \textit{MO} were malicious and did not respect the security of the session keys, a conspiracy attack with the proxies to decrypt all the fragments of the content for the buyer $B_i$ would require that \textbf{all} the proxies (which must be at least two for each purchase) colluded with \textit{MO}. Thus, the minimum coalition required to frame an innocent buyer would be formed by at least three malicious parties: \textit{MO} and two malicious proxies. Forming such coalition would be too risky for \textit{MO}, since one (or more) of the proxies may be honest and refuse to become part of the collusion. Then, the honest proxy (or proxies) could report the collusion between \textit{MO} and the remaining proxies to $T_A$, who would take the necessary actions against \textit{MO} and the involved malicious proxies.
\end{rmk}

% % % % % % % % % % % % % % % % % % % % % % % % % % % % % % % % % % % 
\begin{thm}
	\label{thm_protocol2}
	In Protocol \ref{protocol2}, a malicious proxy trying to obtain the symmetric keys from the database of \textit{MO} would be detected.
\end{thm}

This theorem is already provided and proved in \citep{Me14}.

%\begin{pf}
%To provide resistance against this attack, Protocol \ref{protocol2} inherits security of the four-party anonymous communication protocol proposed in \cite{Me14}. According to the theorem provided in \cite{Me14}, there are two possibilities for a situation in which a malicious proxy tries to obtain the key $K$ by sending the handle $r$ to \textit{MO}. 
%\begin{enumerate}
%\item If a key $K$ has already been retrieved by a child buyer from the transaction database by using the handle $r$, \textit{MO} would either block or remove the register containing $K$. 
%\item The proxy obtains the key $K$ in case the child buyer has not retrieved it yet from \textit{MO}. But in this case, the child buyer will find the corresponding register either blocked or removed. Thus, the child buyer can report this malicious activity of the proxy to the authorities and the \textit{MO}. The child buyer has enough information about a proxy  (such as pseudonym and IP address) that he/she can identify the malicious proxy. Note that \textit{MO} is assumed to be honest for the management of the symmetric keys (Section \ref{sec:assumptions}). 
%\end{enumerate}
%Thus, a malicious proxy trying to obtain $K$ would be detected, since \textit{MO} would block the corresponding register either to the proxy or to the child buyer. This completes the proof. \qed
%\end{pf}
% % % % % % % % % % % % % % % % % % % % % % % % % % % % % % % % % % % 

\begin{thm}
	\label{thm3}
	If there is no collusion, the identity of the illegal re-distributor of the content $X'$ can be recovered in Protocol \ref{protocol3} without requiring the decryption of any of the fingerprints stored in \textit{MO}'s database.
\end{thm}
\begin{pf}
	% On finding a pirated copy, $T_A$, a trusted party of the system (Section \ref{sec:entities}), extracts the fingerprint $f'$ from the traced copy using the fingerprint extraction scheme. Then, $T_A$ permutes $f'$ using $\sigma$, encrypts it with $K_{pA}$ (using also the function $\mathcal{F}$ and the sequence $r_j$) and sends the result to \textit{MO}. Then, \textit{MO} searches the permuted and encrypted fingerprint in its database, which stores the permuted and encrypted fingerprints of all the buyers. Thus, \textit{MO} performs the search in an encrypted domain, and on finding a perfect match, outputs the pseudonym of the illegal re-distributor. The real identity of the traitor can be revealed by \textit{Me}, who has access to the buyers' database. 
	Given the re-distributed copy $X'$ of the content, the basic traitor tracing mechanism (without collusion) is carried out as follows:
	\begin{enumerate}
		\item $X'$ is sent to the tracing authority $T_A$. Then Protocol \ref{protocol3} is run in the following steps.
		\item $T_A$ extracts the fingerprint $f'$ from the content $X'$ using the extraction algorithm as follows: 
		$$f' \leftarrow\mathrm{Extract}(X',\mathrm{parameters}).$$
		\item \label{p3:step3} $T_A$ computes $E(\sigma(f'))$ as per Eq.~\ref{eq:1} (provided in Step \ref{step:eq1} of Protocol \ref{protocol1}), replacing $f_i$ with $f'$. In this step, $T_A$ uses $\sigma$, $K_{pA}$, $\mathcal{F}$ and the sequence $r_j$.
		\item $T_A$ sends $E(\sigma(f'))$ to \textit{MO}.
		\item \label{p3:step5} \textit{MO} searches $E(\sigma(f'))$ in the transaction database that contains $\left\{E(\sigma(f_i))\right\}$ for all the buyers, and returns the pseudonym of the corresponding buyer ($B'$) to $T_A$. As detailed in the assumptions of Section \ref{sec:assumptions}, \textit{MO} cannot cheat in this step.
		\item $T_A$ sends the pseudonym of $B'$ to \textit{Me} and requests the real identity $\mathrm{Id}(B')$ of the buyer corresponding to the pseudonym of $B'$.
		\item \textit{Me} sends the corresponding real identity, $\mathrm{Id}(B')$, retrieved from the buyers database, to $T_A$.
	\end{enumerate}
	As it can be seen, the re-distributor is identified as $\mathrm{Id}(B')$ using only the encrypted and permuted fingerprint $E(\sigma(f'))$ as computed by the tracing authority $T_A$ in Step \ref{p3:step3} above. None of the fingerprints stored in the transaction database of \textit{MO} is decrypted in Step \ref{p3:step5} above, which completes the proof.
	\qed
\end{pf}
% % % % % % % % % % % % % % % % % % % % % % % % % % % % % % % % % % % 

\begin{thm}
	In case of collusion of $c\leq c_0$ buyers, a set of $c$ colluders can be traced efficiently in Protocol \ref{protocol4} using \citeauthor{NuFuHaKiWaOgIm07}'s traitor-tracing algorithm in the encrypted domain.
\end{thm}
\begin{pf}
	%In case a colluded copy is found by $T_A$, it extracts the fingerprint from it, permutes it, encrypts it and sends the permuted and encrypted fingerprint to \textit{MO}, who is responsible for initiating the traitor-tracing algorithm of \citeauthor{NuFuHaKiWaOgIm07}'s codes (Algorithm \ref{code2}). \textit{MO} performs a comparison between the permuted and encrypted bits of the traced and tested fingerprints and sends the output of this comparison to $T_A$, who finally completes the computation of the scores of the buyers. This protocol is carried out without requiring to decrypt any single segment of the (buyers) tested fingerprints (as shown in Algorithm \ref{code3}). The indexes of the buyers whose scores are larger than a given global threshold ($\mathcal{T}$) are then sent to \textit{MO}, who outputs the pseudonyms of the guilty buyers. Finally, \textit{Me} discloses the real identities of the colluders. Thus, the colluders in Protocol \ref{protocol4} can be successfully identified as long as the number of colluders are restricted to $c\leq c_0$. 
	
	Given the re-distributed and pirated copy $Y$, the traitor tracing mechanism (with collusion) is carried out as follows:
	\begin{enumerate}
		\item $Y$ is sent to the tracing authority $T_A$. Then Protocol \ref{protocol4} is run in the following steps.
		\item $T_A$ extracts the fingerprint $f'$ from the pirated content $Y$ using the extraction algorithm as follows: 
		$$f' \leftarrow\mathrm{Extract}(Y,\mathrm{parameters}).$$
		\item \label{p4:step3} $T_A$ computes $E(\sigma(f'))$ as per Eq.~\ref{eq:1} (provided in Step \ref{step:eq1} of Protocol \ref{protocol1}), replacing $f_i$ with $f'$. In this step, $T_A$ uses $\sigma$, $K_{pA}$, $\mathcal{F}$ and the sequence $r_j$. 
		\item $T_A$ sends $E(\sigma(f'))$ to \textit{MO}.
		\item \label{p4:step5} \textit{MO} runs Algorithm \ref{code2}:
		$(\omega_1,\omega_2,D_0,D_1) \leftarrow \text{Algorithm \ref{code2}}(E(\sigma(f')), p,\left\{E(\sigma(f_i))\right\}).$
		The obtained outputs $\omega_1$, $\omega_2$, $D_0$ and $D_1$ are sent to $T_A$.
		\item \label{p4:step6} $T_A$ runs Algorithm \ref{code3}:
		$\left\{S_i\right\} \leftarrow \text{Algorithm \ref{code3}}(\omega_1,\omega_2,D_0,D_1,f',\sigma)$ and obtains the scores for all buyers in the transaction database.
		\item $T_A$ finds $\left\{j:S_j>\mathcal{T}\right\}$ from the set $\left\{S_i\right\}$ (the indexes of the scores greater than the global threshold), and sends the obtained set of indexes $\left\{j\right\}$ to \textit{MO}.
		\item \textit{MO} retrieves, from the transaction database, the pseudonyms of the buyers $\left\{B_j\right\}$ corresponding to the set of indexes $\left\{j\right\}$ obtained in the previous step, and returns them to $T_A$. As detailed in the assumptions of Section \ref{sec:assumptions}, \textit{MO} cannot cheat in this step.
		\item $T_A$ sends the set of pseudonyms of the buyers $\left\{B_j\right\}$ to \textit{Me}, and requests the real identities $\left\{\mathrm{Id}(B_j)\right\}$ of the buyers corresponding to those pseudonyms.
		\item \textit{Me} sends the set of the corresponding real identities, $\left\{\mathrm{Id}(B_j)\right\}$, retrieved from the buyers database, to $T_A$.
	\end{enumerate}
	Hence, the colluders are identified as $\left\{\mathrm{Id}(B_j)\right\}$, provided that the total number of colluders $c$ satisfies $c\leq c_0$. The traitor tracing protocol uses only the encrypted and permuted fingerprint $E(\sigma(f'))$ as computed by the tracing authority $T_A$ in Step \ref{p4:step3} above. None of the fingerprints stored in the transaction database of \textit{MO} ($\left\{E(\sigma(f_i))\right\}$)  is decrypted in Step \ref{p4:step5} above. \citeauthor{NuFuHaKiWaOgIm07}'s traitor-tracing algorithm is carried out in the encrypted domain in Steps \ref{p4:step5} and \ref{p4:step6} provided above (through Algorithms \ref{code2} and \ref{code3}), which completes the proof.
	\qed
\end{pf}

% % % % % % % % % % % % % % % % % % % % % % % % % % % % % % % % % % % 
\subsubsection{Security and privacy attacks on the protocols}
\label{sec:attacks}

This section discusses several attack scenarios presented in Sections \ref{sec:model} and \ref{sec:framing}, which may occur during the execution of the four protocols of the system.
\begin{enumerate}
	\item {\textit{\textbf{Buyer frameproofness:}}} The possible collusion of proxies and \textit{MO} cannot frame an honest buyer and held him/her responsible for illegal re-distribution (formally proved in Theorem $1$). However, other attacks may be possible, namely 1) a malicious \textit{MO} may try to disclose the fingerprint of an honest buyer through the inspection of the fingerprints stored in its database; 2) all the proxies of the same transfer may act maliciously and combine their segments to obtain an honest buyer's fingerprint; 3) a malicious party may try a brute force attack to decrypt the segments to disclose some buyer's fingerprint; and 4) a collusion between \textit{MO} and $T_A$ may try to frame an innocent buyer. These four scenarios are discussed below.
	
	In the first scenario, \textit{MO} cannot reconstruct any buyer's fingerprint directly without the private key of $T_A$. However, for each bit position of the encrypted fingerprints, there are only two possible values (which are different for different positions), which correspond to encrypted binary bits (`0' and `1'). The ciphertext only attack reported in Section \ref{sec:model} would consist of trying to differentiate between encrypted `0's and `1's. Nevertheless, even if \textit{MO} managed to differentiate between all the encrypted `0's and `1's, it would impossible for it to reconstruct any buyer's fingerprint because of the permutation of the fingerprints' bits. Without the permutation key $\sigma$, known only by $T_A$, \textit{MO} would need to compute roughly $l!$ combinations to obtain a correct fingerprint, which would be computationally infeasible for a sufficiently large size of the fingerprint. 
	
	In the second case, the proxies need the secret key of $T_A$ to decrypt each segment of the fingerprint and, hence, they cannot obtain the cleartext of any part of the fingerprint. In addition, they cannot access the decrypted fragments of the contents, as proven in Theorem \ref{thm_protocol2}.
	
	In the third scenario, some malicious party having access to the fragments of the content of the $M$ seed buyers may try a brute force attack to disclose a fingerprint. Each fragment of the content contains a segment embedded into it and, hence, the cleartext segments could be extracted from the fragments if the malicious party had access to the extraction algorithm and keys. Once the cleartext segments are available to the malicious party, they can be encrypted with the public key of $T_A$. If this attacking party also intercepts the encrypted segments sent from the proxies to $T_A$, by inspection, he/she may try to associate the encrypted segments with their cleartext counterparts. However, the segments are not encrypted one by one (in Protocol \ref{protocol1}), but grouped in sets of $m$ contiguous segments (Step 5). As discussed in \citep{Me14}, due to numerical explosion, this prevents such a brute force attack. For example, if $M=10$ and $m=32$, there would be $10^{32}$ possible combinations for each set of $m$ consecutive segments, which would be more than enough for security against this kind of attack.
	
	%However, it is possible that the proxies perform fragment-by-fragment brute force attack to reconstruct a buyer's fingerprint. This attack is infeasible due to the following reasons: 1) each fragment contains $m$ consecutive segments and each segment carries a long fingerprint codeword. For example, if a single segment carries $b=788$-bits of the fingerprint and each bit is encrypted with $1024$-bit key, then a proxy requires $2^{1024}$ time to decrypt each bit. Thus, the decryption of a segment would require $788 \times 2^{1024}$ time, and 2) since there are $m$ fragments and each fragment contains $m$ consecutive segments, decryption of one fragment would require $m \times (b\times 2^{1024})$, e.g. if each fragment contains $10$ segments, then decryption would require $10 \times (788 \times 2^{1024})$, which is computationally infeasible.
	
	Finally, in the last scenario, the collusion can be disregarded since $T_A$ is an entity that is trusted by all the parties of the system (as described in Section \ref{sec:entities}).
	
	\item {\textit{\textbf{Proxy authentication attack:}}} This attack can be resisted by employing an anonymous two-party authenticated key exchange protocol \citep{WaWaXu} in Protocol \ref{protocol2}. When a child buyer selects a proxy peer to obtain the fragments of a content from the parent buyers, he/she can initiate an anonymous two-party authentication protocol with that proxy. After successful authentication, the proxy transfers the fragments of the content from the parent buyers to the child buyer. The details of such anonymous key exchange protocol and its security analysis can be found in \citep{qmr15}. 
	
	\item {\textit{\textbf{Man-in-the middle attack:}}} It is impossible for an attacker to intercept and decrypt the messages between a buyer and a proxy, since he/she would require the session key used for encrypting the content. In order to obtain the session key $K$, the attacker needs to eavesdrop on the communication between \textit{MO} and the child buyer. However, the deployment of PKI in the system (Section \ref{sec:assumptions}) ensures mutual authentication between entities (\textit{MO}, $B_{i}$, proxy) and thus, the communication between the entities is authenticated and the possibility of eavesdropping can be defied.
	
	\item {\textit{\textbf{Database attack:}}} An attacker cannot obtain the fingerprint stored at \textit{MO}'s end, since only \textit{MO} has an access to transaction database. However, it may be possible for the attacker to try to eavesdrop on the communication between $T_A$ and \textit{MO} to obtain the fingerprint. This attack is resisted due to two facts: 1) the fingerprint transferred from $T_A$ to \textit{MO} is permuted with a random permutation key $\sigma$ and encrypted with the public key of $T_A$, and 2) the communication between \textit{MO} and $T_A$ is authenticated due to the existence of a PKI in the system.
	
	\item {\textit{\textbf{Copyright protection:}}} In order to prevent a guilty buyer from repudiating an act of illegal re-distribution, it is necessary that the fingerprint embedded in the buyer's copy and its corresponding permuted and encrypted fingerprint recorded by \textit{MO} be identical. However, there are two possible ways by which the buyer could deny his/her wrongful act, namely 1) if the fingerprint leaks during the execution of the traitor-tracing protocol; and 2) if fake segments are reported by one or more proxies to $T_A$ in Protocol \ref{protocol1}.
	
	The first possibility can be disregarded since $T_A$ is an entity that is trusted by all the parties of the system (as described in Section \ref{sec:assumptions}). Thus, a buyer cannot accuse $T_A$ of leaking the fingerprint in the execution of any protocol. For security against the second scenario, the proposed scheme adopts the same approach as mentioned in \citet{Me14}, i.e. the encrypted fragments of the content and the encrypted embedded segments are both signed. In this way, $T_A$ could randomly request the verification of the signatures of the set of contiguous segments reported by a proxy. If the signature verification failed, the proxy would be accused of malicious behavior. 
	
	\item {\textit{\textbf{Buyer's privacy:}}} In order to protect buyer's privacy from a possible coalition of the transaction monitor and the merchant, or similarly, one of the proxies chosen by the buyer and the merchant, we can use the same approach as the scheme of \citet{Me14}. As discussed in \citep{Me14}, for the first case, the proxies can encrypt the pseudonyms of the buyers with the public key of $T_A$, so that the coalition of \textit{Me} with \textit{MO} cannot break a buyer's privacy. In the second case, the idea of encrypting the buyer's pseudonym with the public key of $T_A$ would not work, since a malicious proxy could decide not to encrypt the pseudonym and form a coalition with \textit{Me} to break the privacy of a buyer. However, \textit{Me} would not be interested in breaking the privacy of his/her clients, since privacy is considered to be one of the clear advantages of the proposed distribution system. Nevertheless, in case \textit{Me} acts maliciously and forms a coalition with a proxy, this possibility can be circumvented by adopting the approach proposed in \citep{qmr15}, where a trusted party, called the certification authority $\mathit{CA_{R}}$, is responsible of issuing certificates to the buyer. $\mathit{CA_{R}}$ will help each buyer in the generation of his/her pseudonym by providing a secret number. The certificate is used to prove that the pseudonym is correctly registered to $\mathit{CA_{R}}$ and only $\mathit{CA_{R}}$ knows about the real identity of the buyer. We can use the same approach here to prevent a malicious \textit{Me} from breaking a buyer's privacy. Instead of having \textit{Me} to manage the buyers' database, $\mathit{CA_{R}}$ could be introduced in the system to authenticate a buyer at time of registration. Once authenticated, $\mathit{CA_{R}}$ would generate a random number and share it with the buyer for the the generation of a pseudonym. Accordingly, the real identity of the guilty buyer would be disclosed by $\mathit{CA_{R}}$ in the traitor-tracing protocols (Protocols \ref{protocol3} and \ref{protocol4}), instead of \textit{Me}. 
	
	Finally, a buyer's privacy could be threatened by the communication between child and parent through a unique proxy. As discussed by \citep{Me14}, this scenario can be evaded by enforcing a child buyer to select a chain of three or more proxies for each set of fragments, such that only two of the proxies of the chain would have access to pseudonyms (either the parents' or the child's).
\end{enumerate}

% % % % % % % % % % % % % % % % % % % % % % % % % % % % % % % % % % % 

\subsection{Theoretical and simulation results for the anti-collusion mechanism}
\label{sec:simulated}
In this section, we present the theoretical and simulation results of the fingerprinting scheme against different types of collusion attacks. 

\subsubsection{Theoretical analysis of the traitor-tracing protocol}
\label{sec:Theoretical}
This section presents a theoretical analysis of the calculation of the fingerprints's scores in the traitor-tracing protocol (Protocol \ref{protocol4}).

In both the standard and the modified versions of \citeauthor{NuFuHaKiWaOgIm07}'s traitor-tracing mechanism,
the (bitwise) score $s^{(k)}_{i}$ of the $i$-th buyer ($B_i$), for $k=1,\dots,l$, is computed using an accusation function $w$, which depends on a discrete bias distribution function $P$, which is a sequence of random values $0<p^{(k)}<1$. The total score $s_i$ of $B_i$ is obtained by taking the average of all bitwise scores for the corresponding fingerprint $f_i$. The values of $w$ decrease as $P$ increases. This implies that the higher the probability of the symbol at this position, the smaller positive amount will be contributed to the total score, and vice versa. According to the marking assumption, a colluded fingerprint consists of the combination of detectable and undetectable positions. The undetectable positions are those for which the embedded bit is the same for all colluders. On the other hand, the bits of the detectable positions are not identical for all colluders. The colluded fingerprint is formed by copying the bit values for all the undetectable positions and ``random values'' (`0's, `1's or erasures) in the detectable positions. The values of the undetectable positions are not really random, since they depend on the specific collusion attack (minimum, maximum or average, for example). The colluded fingerprint is then used as an input in the traitor-tracing protocol (Protocol \ref{protocol4}) to identify a set of colluders. 

%$For example, for four colluders $c_1$, $c_2$, $c_3$ and $c_4$, for an undetectable position $u_{i} \in \left\{0,1\right\}$, where $i \in (1,\ldots,l)$, we have $u^{i}_{c_{1}}=u^{i}_{c_{2}}=u^{i}_{c_{3}}=u^{i}_{c_{4}}$. The detectable positions $d_i \in \left\{0,1\right\}$ for four colluders are those for which the embedded fingerprint bit differ from each other, $d^{i}_{c_{1}}\neq d^{i}_{c_{2}} \vee d^{i}_{c_{1}} \neq d^{i}_{c_{3}} \vee d^{i}_{c_{1}} \neq d^{i}_{c_{4}} \vee d^{i}_{c_{2}} \neq d^{i}_{c_{3}} \vee d^{i}_{c_{2}} \neq d^{i}_{c_{4}} \vee d^{i}_{c_{3}} \neq d^{i}_{c_{4}}$. The colluded fingerprint is then used in the traitor-tracing protocol (Protocol \ref{protocol4}) to trace the colluder (or a set of colluders). 

It can be seen, in Algorithms  \ref{code2} and \ref{code3}, that $s^{(k)}_{i}$ results in a positive value when the (permuted and encrypted) bits of the colluded and tested fingerprints are identical, and negative in case of a bit difference. Thus, it may be inferred from this calculation criterion that the bitwise scores of the colluders must result in all positive values and, similarly, that the bitwise scores of the innocent buyers must result in all negative values. However, this statement cannot not hold true, since the colluded fingerprint is generated assuming the marking assumption, i.e. it consists of undetectable and detectable bits. For colluders, all undetectable bits will provide a positive score with the tested fingerprint, whereas the detectable bits can provide positive or negative scores. Similarly, for non-colluders, some of the bits of the fingerprint can, by chance, be identical to that of the colluded fingerprint, contributing with a positive score. The remaining bits will contribute with a negative score. Thus, in the score computation for a colluder or an innocent buyer, the bitwise scores can result in both positive and negative values. However, for a colluder, a higher number of positive bitwise scores would contribute to its total score (since \citeauthor{NuFuHaKiWaOgIm07}'s traitor-tracing algorithm outputs the guilty buyers with the highest scores). Similarly, the positive values due to the marking assumption can result in a positive score for an innocent buyer, but this score would be comparatively much smaller than that of the colluder (as evident from the simulation results presented in Section \ref{sec:simulation}). Therefore, it can be concluded that, for colluders, the number of positive bitwise scores must be greater than the number of positive bitwise scores for innocent buyers. According to this discussion, a global threshold can be defined to determine whether the codeword belongs to a colluder or an innocent buyer. 

%For example, if all the bits of colluders are $1$, then it is an undetectable bit ($1$), and similarly for all $0$ bits. Thus, when this undetectable colluded bit $1$ is compared with the corresponding bit of the tested fingerprint, which might be $0$, it results in a negative score. Similarly, if the undetectable bit is $0$ and the corresponding bit of the tested fingerprint is $1$, it would result in negative score as well. Thus, the previous statement about bitwise score calculation does not hold true.

%In this regard, bitwise scores are calculated for $1,000$ colluded fingerprints, which are generated by considering collusion attack with $4$ colluders each. In case of colluders, for a fingerprint with length $58,312$, at least $36,000$ bitwise scores are positive, whereas, the maximum number of positive scores for innocent buyers are $29,984$. This result implies that a fingerprint belongs to a colluded fingerprint, if at least $61$\% of its correlation score results in positive values, else the fingerprint belongs to an innocent buyer. 

The above discussion holds when the whole fingerprint is a codeword of \citeauthor{NuFuHaKiWaOgIm07}'s codes. However, in the proposed work, the fingerprint is generated segment-wise by embedding different codewords of \citeauthor{NuFuHaKiWaOgIm07}'s codes in different positions of the fingerprint. The following analysis shows how the traitor-tracing algorithm also works segment-wise, as in the proposed method, since the following features hold:
\begin{enumerate}
	\item Each segment $g_j, j=1,\dots,n_s$ of a fingerprint is a codeword of the \citeauthor{NuFuHaKiWaOgIm07}'s codes resistant against $c \leq c_0$ colluders.
	\item For each segment position, there are $M$ different values, with $M$ being a small number (the number of seed buyers). Hence, the segments will be relatively short codewords\footnote{The same codebook can be used for all segment positions of the fingerprint, or a different codebook can be generated for each position. This makes no difference in the final result.}.
	\item The number of seed buyers must be larger than the collusion resistant capacity of the codes: $M>c_0$ (for example, we can choose $M \geq 2c_0$).
	\item If the number of colluders $c$ satisfies $c\leq c_0$, all the segments $g_j$ of all the colluders will contribute with a high score $\mathcal{S}_j$ such that $T_{\min} \leq \mathcal{S}_j \leq T_{\max}$, where $T_{\min}$ and $T_{\max}$ are some thresholds. The segment score $\mathcal{S}_j$ will be the average of the bitwise scores obtained for all bits in that segment. 
	\item Hence, for a colluder, the score for each segment will be $T_a$ on average, for some $T_a$ such that $T_{\min} \leq T_a \leq T_{\max}$.
	\item The total number of segments available for a position is $M$ and, thus, a non-colluder has  probability $c/M$ of each segment being equal to the same segment of one of the colluders. These segments will contribute with high scores $\mathcal{S}_j$ such that $T_{\min} \leq \mathcal{S}_j \leq T_{\min}$, which are $T_a$ on average. 
	\item On the other hand, for a non-colluder, the remaining segments contribute with a low score $\mathcal{S}_j \leq t_{\max}$, where $t_{\max}$ is a threshold such that $t_{\max} < T_{\min}$. The probability of a segment being different to those of all colluders is $(M-c)/M$. These scores will be typically negative.
\end{enumerate}

Under these conditions, we can make the following approximations:
\begin{enumerate}
	\item The score for all colluders will be above $T_{\min}$ always, and $T_a$ on average, since all of their segments produce such positive scores. 
	\item For non-colluders, there will be, on average:
	\begin{itemize}
		\item $(c/M) \cdot n_s$ segments contributing with scores above $T_{\min}$, and $T_a$ on average. 
		\item $(M-c)/M \cdot n_s$ segments contributing with small scores below $t_{\max}$, usually negative on average.
	\end{itemize}
	Hence, the scores of non-colluders will be typically below $(c/M)T_a \leq (c_0/M) T_a$. If, as suggested, we choose the number of seed buyers such that $M \geq 2c_0$, then we have that the scores of non-colluders will be typically below 
	$$ \frac{c_0}{M} T_a \leq\frac{c_0}{2c_0} T_a =0.5T_a,$$
	and clearly distinguishable from those of colluders, which will be around $T_a$. This is illustrated by means of simulations in the next section.
\end{enumerate}

In this situation we can define a global threshold ($\mathcal{T}$) to distinguish between non-colluders and colluders. Given that $M \geq 2c_0$, a threshold around $0.75 T_a$ will be sufficient to separate colluders (whose scores will be around $T_a$) from non-colluders (whose scores will be typically below $0.5T_a$). Even if a non-colluder happens to provide, by chance, a score somewhat larger that $0.5T_a$, it will still be far from $0.75T_a$. Hence, we can use $\mathcal{T}=0.75T_a$ as a global threshold to discriminate between colluders and innocent buyers.

%%%%%%%%%%%%%Figure 1%%%%%%%%%%%%%%%%%%%%%%%
\begin{figure}[p]
	
	\centering
	\subfloat[Average attack: colluders' set = $\{B_{970},B_{3495},B_{3686},B_{5097}\}$]{ 
		%\graphicspath{G:\PhD Docs\elsarticle-ecrc\Avg.pdf}
		\includegraphics[width=12cm]{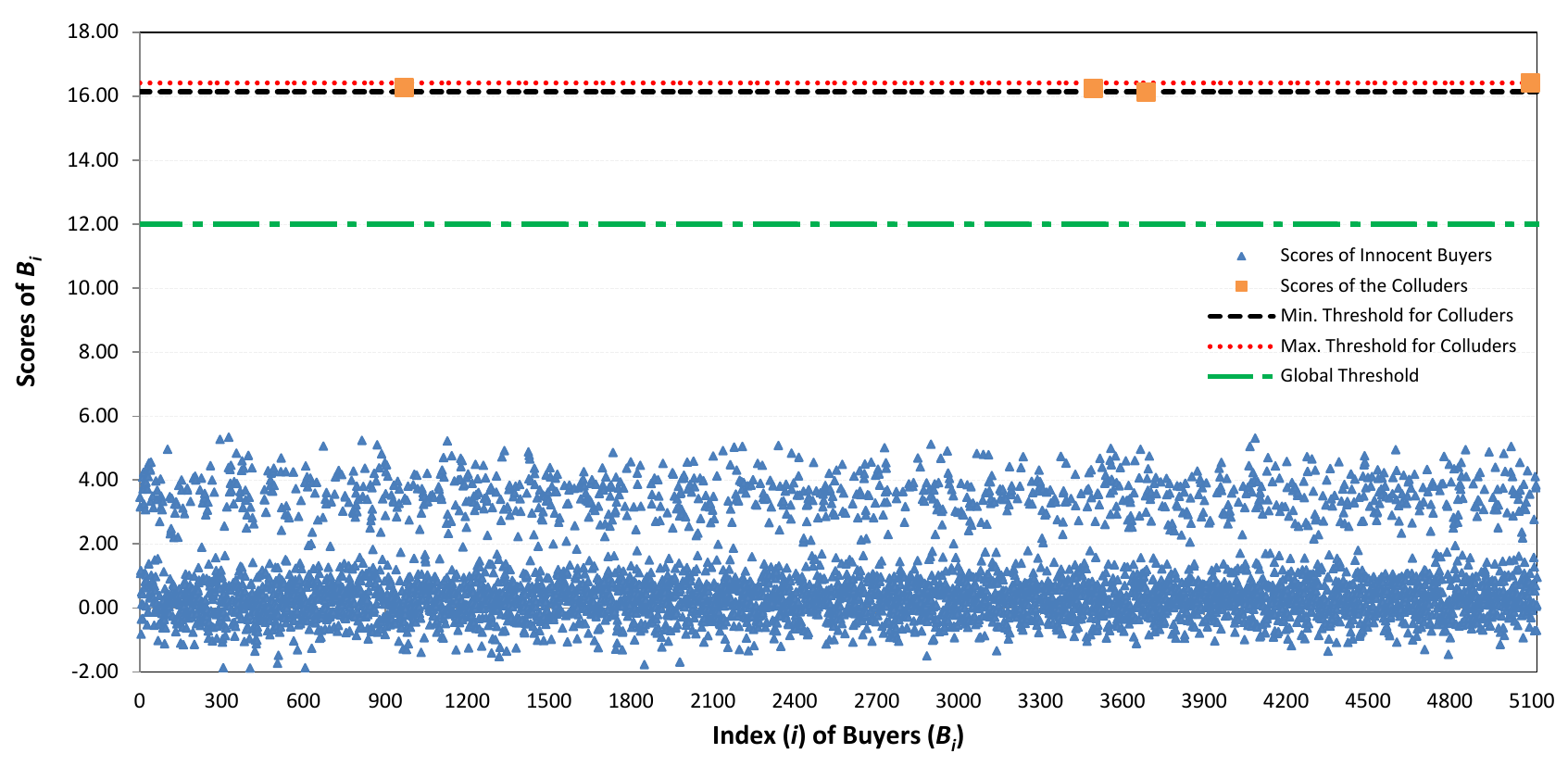}
		\label{fig:fig4}}\\
	\subfloat[Minimum attack: colluders' set = $\{B_{844},B_{1353},B_{2609},B_{3654}\}$]{
		%\graphicspath{G:\PhD Docs\elsarticle-ecrc\Min.pdf}
		\includegraphics[width=12cm]{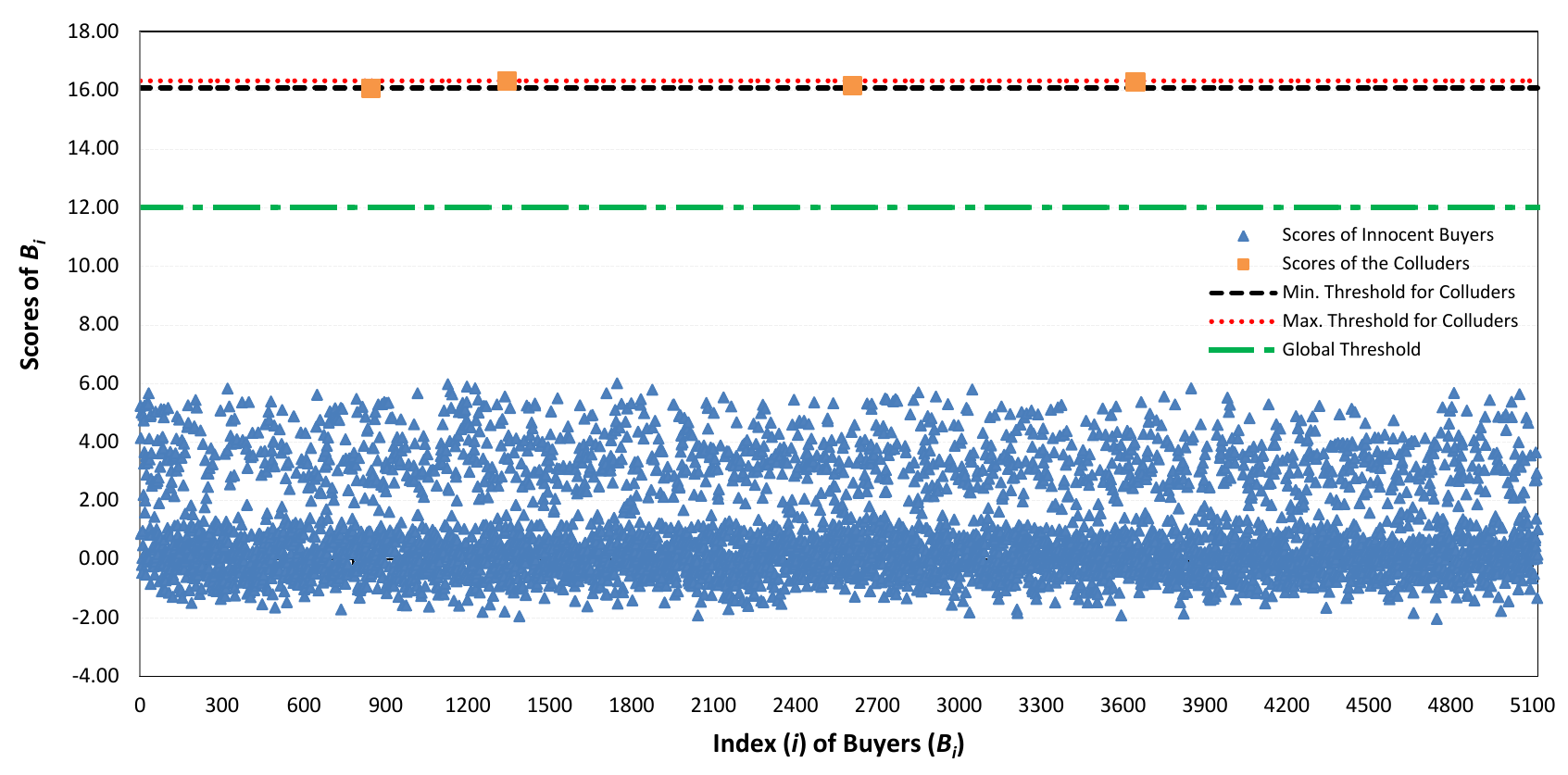}
		\label{fig:fig5}}\\
	\subfloat[Maximum attack: colluders' set = $\{B_{19},B_{1058},B_{1416},B_{4746}\}$]{
		%\graphicspath{G:\PhD Docs\elsarticle-ecrc\Max.pdf}
		\includegraphics[width=12cm]{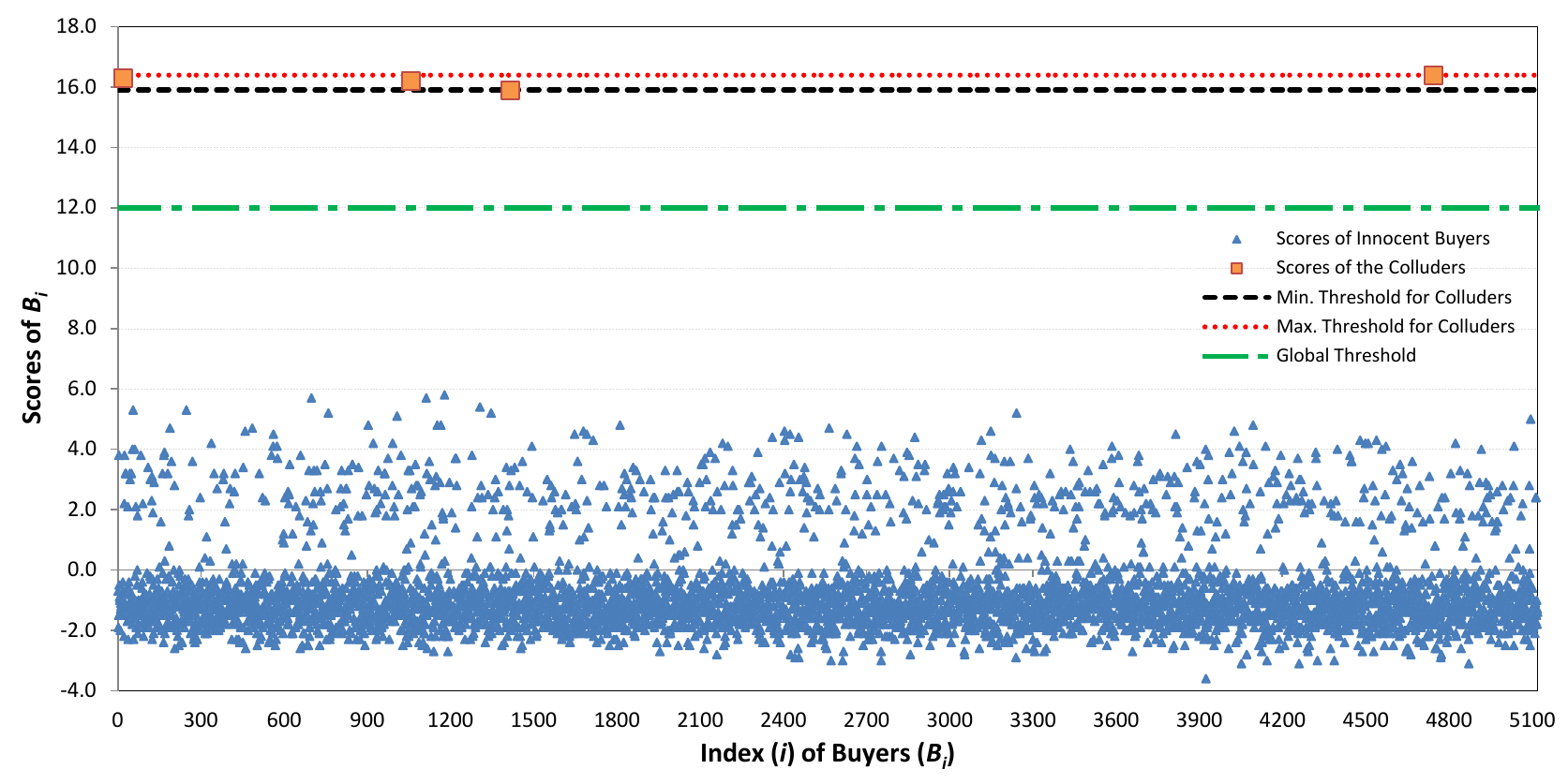}
		\label{fig:fig6}}
	\caption{Scores of buyers for collusion attacks (four colluders)}
	\label{fig:collusion}
\end{figure}
%%%%%%%%%%%%%%%%%%%%%%%%%%%%%%%%%%%%%%%%%%%

\subsubsection{Simulation results}
\label{sec:simulation}
This section presents simulated experiments results to illustrate the security of the proposed fingerprinting scheme against the linear (averaging) and non-linear (minimum and maximum) collusion attacks described in Section \ref{sec:collusion}. 
%The experiments have been carried out in Matlab 7.0 on a workstation equipped with an Intel i-$7$ processor at $3.4$ GHz and $8$ GB of RAM. 
In all the simulations presented in this section, the recombined fingerprints are $58,312$-bit sequences, divided into $74$ segments of $788$ bits each. The number of seed buyers was assumed to be $M=10$. For the seed buyers, the segment codewords were generated using the \citeauthor{NuFuHaKiWaOgIm07}'s fingerprint generation algorithm \citep{NuFuHaKiWaOgIm07} with the following inputs: 10 different codewords for segment (which must be equal to the number of seed buyers), $c_0=4$ and $\varepsilon =10^{-3}$. Including the seed buyers (first generation), $\mathcal{K}$ generations of buyers were produced using an exponential growth approach \citep{MeDo13}. For non-seed buyers, the fingerprints were generated using the recombination approach, choosing between two and four parents per buyer at random from all the previous generations, including the seed buyers. This implies that the average number of parents per non-seed buyer was three.

We have performed the experiments for $\mathcal{K}=10$ generations, which results in a total population of $N=M\cdot2^{\mathcal{K}-1}=10\cdot 2^9=5,120$ buyers (including the seed buyers). The fingerprints stored in \textit{MO}'s database were encrypted using the $1024$-bit Paillier cryptosystem. With these settings, $1,000$ random bit collusions were generated for each attack (average, maximum and minimum) with the number of colluders restricted to $c=4$. For each colluded copy, the traitor-tracing protocol with collusion (detailed in Section \ref{sec:traitor}) was applied. 

In the following, we present the simulations results for these linear and non-linear collusion attacks:
\begin{enumerate}
	\item\textbf{Average attack:} We generated $1,000$ random collusions by taking the average of the corresponding bits of the fingerprints of the four colluders. Then, the traitor-tracing protocol (Protocol \ref{protocol4}) was carried out by $T_A$ and \textit{MO} to output the pseudonyms of the colluders. In the traitor-tracing protocol, each colluded fingerprint was tested with all the buyers' fingerprints stored at \textit{MO}'s end, and the corresponding scores were calculated in the encrypted domain. The protocol provided the pseudonyms of the colluders as the buyers with the highest scores, whereas innocent buyers yielded much lower scores. In all $1,000$ cases, the four colluders were successfully identified and no innocent buyer was wrongfully accused.
	
	The collusion resistance of Protocol \ref{protocol4} under the average attack is illustrated in Figure \ref{fig:fig4} for one of those $1,000$ simulations. As it can be seen, the four colluders can be easily identified with a score in the range $[16.28,16.42]$, whereas the remaining $5,116$ buyers provided scores in the range $[-2.00,5.30]$. Since the score of the colluders was around 16 (in fact higher), we could safely set the global threshold to be $\mathcal{T}=0.75\cdot 16=12$ to discriminate between colluders and innocent buyers with no false positives or negatives.
	
	\item\textbf{Minimum attack:} Another $1,000$ simulations were carried out for the minimum attack with analogous results. Figure \ref{fig:fig5} illustrates the collusion resistance of the fingerprinting scheme against the minimum attack for one of those $1,000$ simulations. Through the traitor-tracing protocol, the buyers with scores between $[16.08,16.28]$ were identified as colluders, whereas, the buyers with scores in the range $[-2.04,5.97]$ were output as innocent buyers. As in the averaging attack, the minimum attack yielded 100\% colluders detection rate with 0\% false accusation rate. The same global threshold $\mathcal{T}=12$ was used.
	
	\item\textbf{Maximum attack:} Again, under this attack, $1,000$ random colluded fingerprints were formed by taking the maximum of the corresponding bits of the fingerprints of four randomly selected colluders, with the same performance as in the previous collusion attacks. Figure \ref{fig:fig6} shows the results obtained for one of those $1,000$ simulations, and it can be seen that all four colluders were successfully identified with a score in $[15.90,16.40]$, whereas the innocent buyers provided scores in the range $[-4.00,5.80]$. The same global threshold used in the other two collusion attacks, $\mathcal{T}=12$, provided the correct identification of all colluders with no false positives or negatives.
\end{enumerate}

% % % % % % % % % % % % % % % % % % % % % % % % % % % % % % % 
\begin{figure}[ht]
	\centering
	%\graphicspath{G:\PhD Docs\elsarticle-ecrc\Scores.pdf}
	\includegraphics[width=14cm]{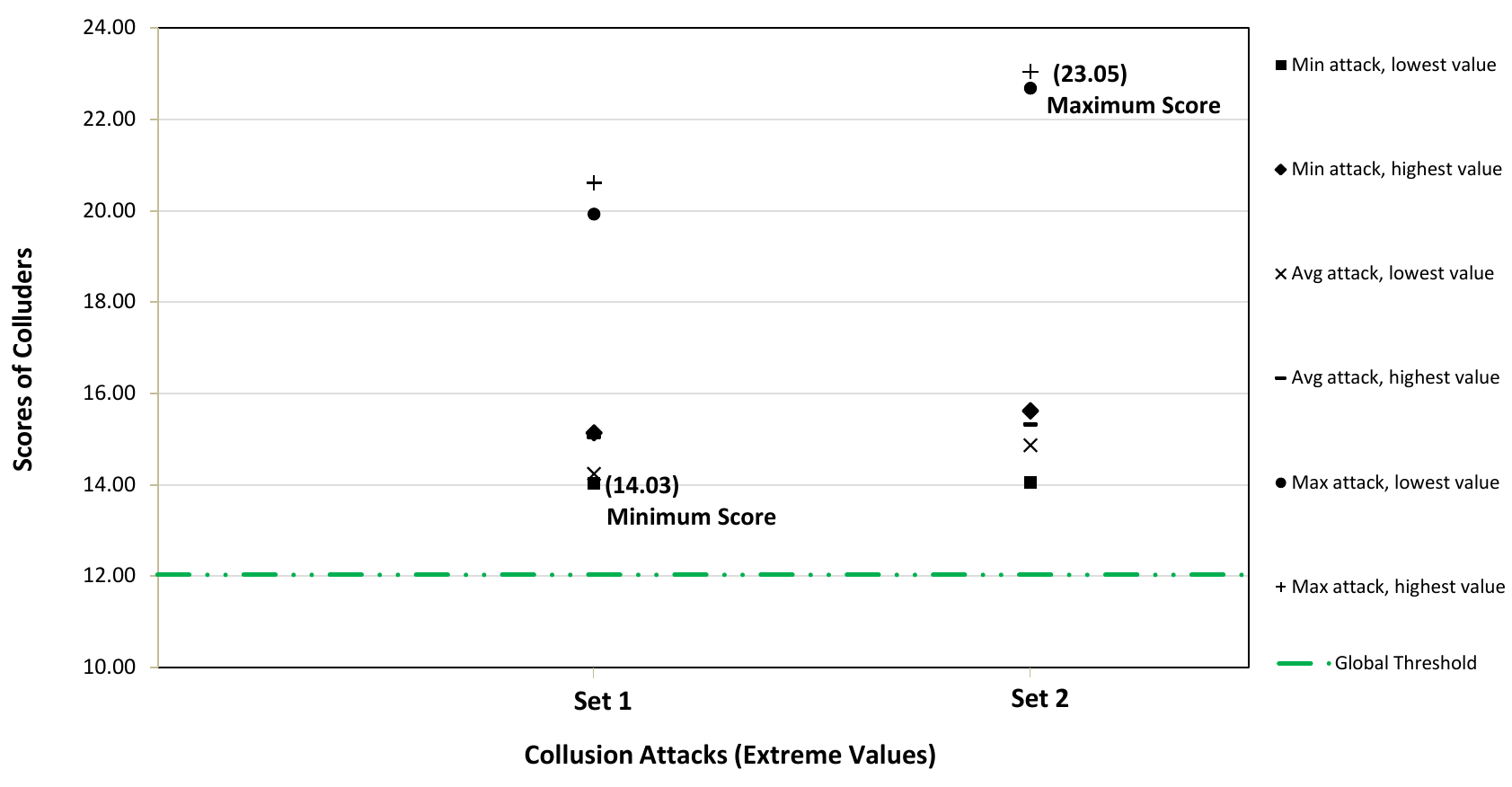}
	\caption{Extreme scores obtained for the colluders}
	\label{fig:fig7}
\end{figure}
%%%%%%%%%%%%%%%%%%%%%%%%%%%%%%%%%%%%%%%%%%%

In Figure \ref{fig:collusion}, it can be seen that $T_{\min}$ and $T_{\max}$ vary in different experiments. Values around 16 for the scores of colluders are usual, but some other values were obtained in some simulations. The range for colluders' scores considering the three attacks in the $1,000$ simulations was $[14.03,23.05]$. Similarly, the scores for non-colluders varied in different simulations, but they were always in the range  $[-4.00,5.97]$. This confirms that the global threshold $\mathcal{T}=12$ works for the $1,000\times 3$ simulated collusions with no false positives or negatives. The scores of the two sets of colluders which lead to the maximum ($23.05$) and minimum ($14.03$) values obtained after $1,000$ simulations are depicted in Figure \ref{fig:fig7}.

It is worth pointing out that more than four colluders could be identified through Protocol \ref{protocol4} if the fingerprint were generated with $c_0>4$ and the number of seed buyers was adjusted accordingly, so as to satisfy $M\geq 2c_0$. However, the larger the collusion resistance of the code, the longer the codewords become and, hence, the longer the fingerprint results. An increase in the length $l$ of the fingerprint requires embedding more bits in the content and the embedding algorithm must provide more capacity. Embedding more bits will have an effect on the quality of the content since, typically, the perceptual quality decreases when capacity increases. Thus, the value $c_0$ shall be selected taking into account the trade-off between collusion resistance and perceptual quality.

\subsection{Performance analysis}
\label{sec:performance}

This section presents a performance analysis of the proposed system in terms of the computational (especially cryptographic) effort required by the entities and the communication cost. To show the performance of the system, the experiments were carried out in Matlab $7.0$ and Java on the audio file ``LoopyMusic", on a workstation equipped with an Intel i-7 processor at $3.4$ GHz and $8$ GB of RAM. The distribution protocol is implemented in Matlab $7.0$, except the public-key ($1024$-bit Paillier) encryption  of the content that is implemented in the Java programming language. The details of the ``LoopyMusic" audio file are presented in Table \ref{tab:5.1}.

%%%%%%%%%%%%%%%%%%%%%%%%%%%%%%%%%%%%%%%%%%%%%%%%%%%%%%%%%%%
\begin{table}[htbp]
	\caption{Details of ``LoopyMusic"}
	\label{tab:5.1}
	\centering
	\begin{tabular}{|l|c|}\hline
		\textbf{Details} & \textbf{Loopy Music}  \\ \hline
		Time Length (min:sec) & $00$:$10$  \\ \hline
		File Size (MB) & $0.89$   \\ \hline
		Format           & WAV        \\ \hline
		Bits per Sample  & $16$    \\ \hline
		Sample Rate (Hz)  & $44100$   \\ \hline
		Channel Mode     & Mono    \\  \hline
	\end{tabular}
\end{table}
% % % % % % % % % % % % % % % % % % % % % % % % % % % % % % % % % %

The bootstrapping of the proposed system is initialized by \textit{Me}, who provides the multimedia content (``LoopyMusic") to the \textit{MO} to carry out the remaining part of the bootstrapping process. \textit{MO} is responsible for the the generation of $M=10$ different \citet{NuFuHaKiWaOgIm07}'s $c$-secure codes and $M=10$ seed copies of ``LoopyMusic". In each seed copy, the fingerprint codes of length $l=58,312$ bits that provide resistance against $c=4$ colluders, are embedded into $74$ segments (with a codeword of $788$ bits per segment) of the audio content using the robust, secure and transparent audio watermarking algorithm proposed by \citet{XiWaZhXuYa13}. In addition, for each seed copy, \textit{MO} encrypts each segment embedded into the fragments of ``LoopyMusic" using the public key of the tracing authority. Table \ref{tab:5.3} presents the overhead of \textit{MO} for bootstrapping of the system.\\

%%%%%%%%%%%%%%%%%%%%%%%%%%%%%%%%%%%%%%%%%%%%%%%%%%%%%%%%%%%
% % % % % % % % % % % % % % % % % % % % % % % % % 
\begin{table}[htbp]
	\caption{Bootstrapping overhead}
	\label{tab:5.3}
	\centering
	\small\addtolength{\tabcolsep}{-2.0pt}
	\begin{tabular}{|c|c|c|c|c|c|} \hline
		\multirow{3}{*}{\textbf{\begin{tabular}[c]{@{}c@{}}File\\Name\end{tabular}}} & \multicolumn{5}{c|}{\textbf{CPU Time (secs)}} \\ \cline{2-6} 
		& \multirow{2}{*}{\textbf{\begin{tabular}[c]{@{}c@{}}Fingerprint\\generation\end{tabular}}} & \multirow{2}{*}{\textbf{\begin{tabular}[c]{@{}c@{}}Embedding\\ (Generation of seed copies)\end{tabular}}} & \multirow{2}{*}{\textbf{\begin{tabular}[c]{@{}c@{}}Communication with\\seed buyers\end{tabular}}} &
		\multirow{2}{*}{\textbf{\begin{tabular}[c]{@{}c@{}}Public-key\\encryption of segments\end{tabular}}} &
		\multirow{2}{*}{\textbf{\begin{tabular}[c]{@{}c@{}}Total\\Time\end{tabular}}} \\
		&  &  &  & & \\ \hline
		LoopyMusic  &  $0.33$  &   $1.65$  &  $0.05$   &  $8.59$ & $10.62$ \\ \hline
	\end{tabular}
\end{table}
% % % % % % % % % % % % % % % % % % % % % % % % % % % % % % % % % %

The overhead of \textit{MO} for the generation of $10$ seed copies is one-time cost only (it could be performed by \textit{MO} in an offline mode). Once the system is bootstrapped, the child buyers obtain their fingerprinted copies of the content by recombining fragments of the content received from the parent buyers. From Table \ref{tab:5.3}, it can be seen that $1024$-bit public-key encryption of the segments contributes to most of the computational costs. It is pertinent to mention that the only overhead (a few seconds depending on the size of the file) for the \textit{Me} is the transfer of the original content to \textit{MO} in the bootstrapping process.\\

Tables \ref{tab:5.4}, \ref{tab:5.5}, \ref{tab:5.6} and \ref{tab:5.7} show the overhead costs calculated for each party (parent buyer, \textit{MO}, proxy and child buyer) involved in the anonymous distribution of the fragments of ``LoopyMusic" between parent and child buyers. We have calculated the costs for a scenario where a child buyer requests a single copy of ``LoopyMusic" from the merchant. Five proxies are used to transfer the fragments of the content from at least three parent buyers (minimum requirement is two parent buyers).

% % % % % % % % % % % % % % % % % % % % % % % % % 
\begin{table}[htbp]
	\caption{Overhead costs of a parent buyer}
	\label{tab:5.4}
	\centering
	\small\addtolength{\tabcolsep}{-2.0pt}
	\begin{tabular}{|c|c|c|c|c|c|} \hline
		\multirow{3}{*}{\textbf{\begin{tabular}[c]{@{}c@{}}File\\Name\end{tabular}}} & \multicolumn{5}{c|}{\textbf{CPU Time (secs)}} \\ \cline{2-6} 
		& \multirow{2}{*}{\textbf{\begin{tabular}[c]{@{}c@{}}Session key and\\ Random number generation\end{tabular}}} & \multirow{2}{*}{\textbf{\begin{tabular}[c]{@{}c@{}}Communication with\\\textit{MO}\end{tabular}}} & \multirow{2}{*}{\textbf{\begin{tabular}[c]{@{}c@{}}Symmetric encryption\\of fragments\end{tabular}}} &
		\multirow{2}{*}{\textbf{\begin{tabular}[c]{@{}c@{}}Communication with\\proxies\end{tabular}}} &
		\multirow{2}{*}{\textbf{\begin{tabular}[c]{@{}c@{}}Total\\Time\end{tabular}}} \\
		&  &  &  & & \\ \hline
		LoopyMusic  &  $0.02$  &   $0.003$  &  $3.40$   &  $0.30$ & $3.72$ \\ \hline
	\end{tabular}
\end{table}
% % % % % % % % % % % % % % % % % % % % % % % % % % % % % % % % % %

Table \ref{tab:5.4} presents the overhead costs of a parent buyer equal to $3.72$ seconds, which includes session key ($K$) and random number ($r$) generation, communication with \textit{MO} --transfer of ($r,K$)--, symmetric encryption of the fragments requested by each proxy using $K$, and communication with each proxy (transfer of $r$ and encrypted fragments). All three parent buyers involved in anonymous content transfer execute the assigned functions in parallel without interfering with each other and, thus resulting in the total overhead equal to $3.72$ seconds instead of $11.16$ seconds.

% % % % % % % % % % % % % % % % % % % % % % % % % 
\begin{table}[htbp]
	\caption{Overhead costs of \textit{MO}}
	\label{tab:5.5}
	\centering
	\small\addtolength{\tabcolsep}{-2.0pt}
	\begin{tabular}{|c|c|c|c|c|} \hline
		\multicolumn{2}{|c|}{\multirow{2}{*}{\textbf{\begin{tabular}[c]{@{}c@{}}File\\Name\end{tabular}}}} & \multicolumn{3}{c|}{\textbf{CPU Time (secs)}} \\ \cline{3-5} 
		\multicolumn{2}{|c|}{} & \multicolumn{3}{c|}{\textbf{\begin{tabular}[c]{@{}c@{}}Communication with\\child buyer\end{tabular}} }\\ \hline
		\multicolumn{2}{|c}{LoopyMusic}&\multicolumn{3}{|c|}{$0.09$} \\ \hline
	\end{tabular}
\end{table}
% % % % % % % % % % % % % % % % % % % % % % % % % % % % % % % % % %

In Table \ref{tab:5.5}, we can see that \textit{MO} takes fraction of a second to send the corresponding $K$ to the child buyer.\\

% % % % % % % % % % % % % % % % % % % % % % % % % 
\begin{table}[htbp]
	\caption{Overhead costs of a proxy}
	\label{tab:5.6}
	\centering
	\small\addtolength{\tabcolsep}{-2.0pt}
	\begin{tabular}{|c|c|c|c|c|c|} \hline
		\multirow{3}{*}{\textbf{\begin{tabular}[c]{@{}c@{}}File\\Name\end{tabular}}} & \multicolumn{5}{c|}{\textbf{CPU Time (secs)}} \\ \cline{2-6} 
		& \multirow{2}{*}{\textbf{\begin{tabular}[c]{@{}c@{}}Communication with\\child buyer\end{tabular}}} & \multirow{2}{*}{\textbf{\begin{tabular}[c]{@{}c@{}}Concatenation of\\encrypted fragments\end{tabular}}} & \multirow{2}{*}{\textbf{\begin{tabular}[c]{@{}c@{}}Public-key encryption\\of concatenated fragments\end{tabular}}} &
		\multirow{2}{*}{\textbf{\begin{tabular}[c]{@{}c@{}}Communication with\\$T_A$\end{tabular}}} &
		\multirow{2}{*}{\textbf{\begin{tabular}[c]{@{}c@{}}Total\\Time\end{tabular}}} \\
		&  &  &  & & \\ \hline
		LoopyMusic  &  $0.30$  &   $0.03$  &  $0.65$   &  $0.05$ & $1.03$ \\ \hline
	\end{tabular}
\end{table}
% % % % % % % % % % % % % % % % % % % % % % % % % % % % % % % % % %

The proxies are solely responsible for transferring the encrypted fragments from the parent buyer to the child buyer in the anonymous distribution protocol. Table \ref{tab:5.6} presents the overhead cost of five proxies equal to $1.03$ seconds that include communication with the child buyer (transfer of $r$ and encrypted fragments), concatenation of the encrypted fragments, encryption of the concatenated fragments with the public key of $T_A$, and transfer of the encrypted fragments to $T_A$. Similar to parent buyers, all proxies execute the assigned functions in parallel, thus reducing the overall distribution time. Thus, it can be easily seen that the overhead contributed by proxies in the distribution protocol is small.
% % % % % % % % % % % % % % % % % % % % % % % % % 
\begin{table}[H]
	\caption{Overhead costs of a child buyer}
	\label{tab:5.7}
	\centering
	\small\addtolength{\tabcolsep}{-2.0pt}
	\begin{tabular}{|c|c|c|c|} \hline
		\multirow{3}{*}{\textbf{\begin{tabular}[c]{@{}c@{}}File\\Name\end{tabular}}} & \multicolumn{3}{c|}{\textbf{CPU Time (secs)}} \\ \cline{2-4} 
		& \multirow{2}{*}{\textbf{\begin{tabular}[c]{@{}c@{}}Communication with \\ \textit{MO}\end{tabular}}} & \multirow{2}{*}{\textbf{\begin{tabular}[c]{@{}c@{}}Symmetric decryption\\of content\end{tabular}}} & 
		\multirow{2}{*}{\textbf{\begin{tabular}[c]{@{}c@{}}Total\\Time\end{tabular}}} \\
		&  &  &   \\ \hline
		LoopyMusic  &  $0.001$  &   $1.10$ & $1.10$ \\ \hline
	\end{tabular}
\end{table}
% % % % % % % % % % % % % % % % % % % % % % % % % % % % % % % % % %

In the distribution protocol, a child buyer performs two tasks, i.e. communication with \textit{MO} (transfer of $r$ to \textit{MO} in order to receive the corresponding $K$) and decryption of the received encrypted fragments with $K$. From Table \ref{tab:5.7}, it can be seen that it hardly takes a second for a child buyer to decrypt the received fragments.\\

From the overheads presented in Tables \ref{tab:5.4}, \ref{tab:5.5}, \ref{tab:5.6} and \ref{tab:5.7}, it can be concluded that the anonymous content transfer between parent buyers and a child buyer is executed in $5.94$ seconds ($3.72+0.09+1.03+1.10$). This is the overall time required for transferring the file for each buyer.
%%%%%%%%%%%%%%%%%%%%%%%%%%%%%%%%%%%%%%%%%%%%%%%%%%%%%%%%%%%%%%%%%%%%5

\subsubsection{Comparative analysis with \citet{qmr16}}
\label{sec:compar}

This section carries out a comparative analysis of the proposed system with PSUM presented in \citep{qmr16}. The comparison focuses on the performance of the systems in terms of computational and communication costs. The reason for selecting PSUM for comparison is the fact that PSUM is more efficient than the systems proposed by \citet{MeDo13a,Me14} and \citet{qmr15} (as evident from the analysis presented in Table \ref{tab:comparative} in Section \ref{sec:comparative}).\\

Table \ref{tab:5.8} shows the computation time of bootstrapping the system using the ``LoopyMusic" audio file for the proposed system and PSUM \citep{qmr16}.

%%%%%%%%%%%%%%%%%%%%%%%%%%%%%%%%%%%%%%%%%%%%%%%%%%%%%%%%%%
\begin{table}[H]
	\caption{Comparison of bootstrapping time}
	\label{tab:5.8}
	\centering
	\small\addtolength{\tabcolsep}{-2.0pt}
	\begin{tabular}{|c|c|c|c|c|c|c|}
		\hline
		\multicolumn{2}{|c|}{\multirow{4}{*}{\textbf{\begin{tabular}[c]{@{}c@{}}File \\ Name\end{tabular}}}} & \multicolumn{5}{c|}{\textbf{Bootstrapping performed by \textit{MO}}} \\ \cline{3-7} 
		\multicolumn{2}{|c|}{} & \multicolumn{5}{c|}{\textbf{CPU Time in secs for the proposed system}} \\ \cline{3-7} 
		\multicolumn{2}{|c|}{} & \multicolumn{5}{c|}{\textbf{Functions}} \\ \cline{3-7} 
		\multicolumn{2}{|c|}{} & \begin{tabular}[c]{@{}c@{}}Fingerprint\\ generation\end{tabular} & Embedding & \begin{tabular}[c]{@{}c@{}}Communication with\\ seed buyers\end{tabular} & \multicolumn{1}{c|}{\begin{tabular}[c]{@{}l@{}}Public-key\\ encryption\end{tabular}} & \begin{tabular}[c]{@{}l@{}}Total \\ Time\end{tabular} \\ \hline
		\multicolumn{2}{|c|}{LoopyMusic} & $0.33$ & $1.65$ & $0.05$ & $8.59$ & \multicolumn{1}{c|}{$10.62$} \\ \hline
		\multicolumn{2}{|c|}{\multirow{4}{*}{\textbf{\begin{tabular}[c]{@{}c@{}}File\\ Name\end{tabular}}}} & \multicolumn{5}{c|}{\textbf{Bootstrapping performed by \textit{Me}}} \\ \cline{3-7} 
		\multicolumn{2}{|c|}{} & \multicolumn{5}{c|}{\textbf{CPU Time in secs for PSUM}} \\ \cline{3-7} 
		\multicolumn{2}{|c|}{} & \multicolumn{5}{c|}{\textbf{Function}} \\ \cline{3-7} 
		\multicolumn{2}{|c|}{} & \multicolumn{4}{c|}{\textbf{DWT+Embedding}}  & \begin{tabular}[c]{@{}l@{}}Total\\ Time\end{tabular} \\ \hline
		\multicolumn{2}{|l|}{LoopyMusic} & \multicolumn{4}{c|}{$0.19$}  & \multicolumn{1}{c|}{$0.19$} \\ \hline
	\end{tabular}
\end{table}
%%%%%%%%%%%%%%%%%%%%%%%%%%%%%%%%%%%%%%%%%%%%%%%%%%%%%%%%%%%%%%%%%55

The last column in Table \ref{tab:5.8} contains the total CPU time of bootstrapping process for both the systems. Please note that bootstrapping is carried out by the monitor in the proposed system, but by the merchant in PSUM. It can be seen that the computational cost of PSUM is comparatively shorter than that of the proposed system. In the proposed system, the public-key encryption of the content is the most expensive operation. However, it must be taken into account that bootstrapping only occurs once in the proposed system, before the first transaction to purchase the content.

%%%%%%%%%%%%%%%%%%%%%%%%%%%%%%%%%%%%%%%%%%%%%%%%%%%%%%%%%%
\begin{table}[H]
	\caption{Comparison of communication time}
	\label{tab:5.9}
	\centering
	\small\addtolength{\tabcolsep}{-2.0pt}
	\begin{tabular}{|c|c|c|c|c|c|c|}
		\hline
		\multicolumn{2}{|c|}{\multirow{4}{*}{\textbf{\begin{tabular}[c]{@{}c@{}}File \\ Name\end{tabular}}}} & \multicolumn{5}{c|}{\textbf{Anonymous content transfer between parent and child buyers}} \\ \cline{3-7} 
		\multicolumn{2}{|c|}{} & \multicolumn{5}{c|}{\textbf{CPU Time in secs for the proposed system}} \\ \cline{3-7} 
		\multicolumn{2}{|c|}{} & \multicolumn{5}{c|}{\textbf{Functions}} \\ \cline{3-7} 
		\multicolumn{2}{|c|}{} & \begin{tabular}[c]{@{}c@{}}Overhead of\\ \textit{Me}\end{tabular} &
		\begin{tabular}[c]{@{}c@{}} Overhead of buyer\\(parent+child)\end{tabular}& \begin{tabular}[c]{@{}c@{}}Overhead of \\ \textit{MO}\end{tabular} & \begin{tabular}[c]{@{}c@{}}Overhead of\\ proxy\end{tabular} & \begin{tabular}[c]{@{}l@{}}Total \\ Time\end{tabular} \\ \hline
		\multicolumn{2}{|c|}{LoopyMusic} & $0.00$ & $4.82$ & $0.09$ & $1.03$ & \multicolumn{1}{c|}{$5.94$} \\ \hline
		\multicolumn{2}{|c|}{\multirow{4}{*}{\textbf{\begin{tabular}[c]{@{}c@{}}File\\ Name\end{tabular}}}} & \multicolumn{5}{c|}{\textbf{Anonymous Content transfer between a merchant and a buyer}} \\ \cline{3-7} 
		\multicolumn{2}{|c|}{} & \multicolumn{5}{c|}{\textbf{CPU Time in secs for PSUM}} \\ \cline{3-7} 
		\multicolumn{2}{|c|}{} & \multicolumn{5}{c|}{\textbf{Functions}} \\ \cline{3-7} 
		\multicolumn{2}{|c|}{} & \begin{tabular}[c]{@{}c@{}}Overhead of\\ \textit{Me}\end{tabular}&\begin{tabular}[c]{@{}c@{}}Overhead of\\ a buyer\end{tabular} & \begin{tabular}[c]{@{}c@{}}Overhead of\\ \textit{MO}\end{tabular} & \begin{tabular}[c]{@{}c@{}}Overhead of\\ Proxy\end{tabular} & \begin{tabular}[c]{@{}l@{}}Total\\ Time\end{tabular} \\ \hline
		\multicolumn{2}{|l|}{LoopyMusic} &$2.19$ &$3.35$ & $6.33$ & $2.51$ & \multicolumn{1}{c|}{$14.38$} \\ \hline
	\end{tabular}
\end{table}
%%%%%%%%%%%%%%%%%%%%%%%%%%%%%%%%%%%%%%%%%%%%%%%%%%%%%%%%%%%%%%%%

In Table \ref{tab:5.9}, we compare the communication time of proposed system for ``LoopyMusic" with the published response time results of \citet{qmr16}'s PSUM. In PSUM, the communication time is calculated as the time taken in base file (a small size file in which approximation coefficients are embedded with the fingerprint) distribution from the merchant to the buyer through proxy peers in the presence of the \textit{MO}, whereas, in the proposed system, the communication time is calculated as the time taken in file distribution from at least three parent buyers to the child buyer through proxy peers. 

In PSUM, the overhead cost of \textit{Me} ($2.19$ seconds) includes decryption of the session and permutation keys, permutation of the approximation coefficients and symmetric encryption of the permuted coefficients. This overhead cost is linear to the number of file requests received from the buyer. If \textit{Me} receives multiple requests of the same content from different buyers, the overhead cost would increase linearly. In the proposed system, the overhead of \textit{Me} is zero since it is not involved in the file distribution protocol, thus providing \textbf{the most practical and efficient content distribution solution to the merchant}. The overhead of \textit{Me} is zero in the proposed system since it is not involved in the file distribution protocol (\textit{Me} participates only in the bootstrapping process) between a parent and a child buyer. The overhead cost of a buyer ($3.35$ seconds) in PSUM includes generation of an anonymous key pair, a certificate, permutation and session keys, and file reconstruction (symmetric decryption, inverse permutation and DWT, and conversion to original audio format). On the other hand, in the proposed system, the overhead costs of a parent and a child buyer contributes to the overall buyer overhead, in which the child buyer performs its tasks in $1.10$ seconds and the parent buyer executes its operations in $3.72$ seconds. In PSUM, the increased overhead of \textit{MO} is due to the fact it is responsible for generation of $10^4$ $c$-secure codewords (which takes $6.01$ seconds and contributes highly to the overall overhead), fingerprint segmentation and permutation ($0.02$ seconds) and communication with the merchant, the buyer and the proxies ($0.30$ seconds). The overhead of the \textit{MO} is minimal ($0.09$ seconds) in the proposed system. For the proxy peers of PSUM, the overhead cost is equal to $2.51$ seconds, which includes transferring of the encrypted and permuted coefficients from the merchant to the buyer, while, the proxy peers in the proposed system takes hardly a second to perform the assigned functions (as shown in Table \ref{tab:5.6}).

It can be seen from the last column of Table \ref{tab:5.9}, that the total file distribution time in the proposed system ($5.94$ seconds) is comparatively shorter than the distribution time using the scheme of \citet{qmr16} ($14.38$ seconds). Thus, it is evident, from the results shown in Table \ref{tab:5.9}, that the proposed distribution protocol of the proposed system permits to \textbf{reduce the communicational burdens of secure fingerprinted content distribution}. 

%%%%%%%%%%%%%%%%%%%%%%%%%%%%%%%%%%%%%%%%%%%%%%%%%%%%%%%%%%%%%%%%5

\section{Conclusions}
\label{sec:conclusions}

This paper presents a novel P2P content distribution system with copyright and privacy protection to the merchant and the end users, respectively. The proposed system exploits the recombined fingerprints idea suggested in \citep{MeDo13,Me14}, which was inspired from the DNA sequences of the living beings. This strategy, which resembles the approach taken in genetic algorithms, is exploited here for multimedia distribution. However, the existing systems based on the recombination approach exhibit some shortcomings that make them inefficient and impractical under certain circumstances.
To overcome the drawbacks of prior art, the first contribution of this paper is the use of state-of-the-art  \citeauthor{NuFuHaKiWaOgIm07}'s (\citeyear{NuFuHaKiWaOgIm07}) codes  \textbf{at segment level}. \citeauthor{NuFuHaKiWaOgIm07}'s codes are shown to work in segment-wise fashion for the first time, in such a way that each segment of the fingerprint is a codeword. The small number of seed buyers makes it possible to obtain short codewords for each segment, resulting in remarkably short fingerprints, and prevents the two-layer encoding required by the previous recombination-based proposals. Also, the proxies in the proposed system do not need to communicate with each other in the presence of the monitor to construct a valid fingerprint at hash level, since the requirement for a hash of the fingerprint has been completely removed in the proposed system. In addition, encryption is applied to the permuted bits of the fingerprint using the public key of the authority, which is a trusted party. For the secure identification of the colluder(s), the second contribution of the paper is a novel traitor-tracing system that performs \textbf{tracing in the encrypted domain}. The proposed traitor-tracing protocol does not require decrypting any single segment of the fingerprint and thus, preserves buyer frameproofness. The security analysis, the simulation results and the performance evaluation show that the proposed system provides a secure, scalable, privacy-preserving, collusion-resistant, anonymous and efficient content distribution mechanism for P2P systems with low computation costs.

Future research should be directed to (1) a real-world implementation of the proposed system for a real P2P application, e.g. as a plug-in that enables the copyright and privacy protection features in a P2P system; and (2) the measurement of some metrics at the application layer, such as bandwidth and latency, with the real-world implementation, in order to evaluate the impact of the proposed content distribution protocol on the user experience.

% % % % % % % % % % % % % % % % % % % % % % % % % % % % % % % % % % % % % % % % % % % % % % % % % % % % % % % %
\section*{Acknowledgment}                             
This work was partly funded by the Spanish Government through grants TIN2011-27076-C03-02 ``CO-PRIVACY'' and TIN2014-57364-C2-2-R ``SMARTGLACIS''.

\bibliographystyle{elsarticle-harv}
\bibliography{drafteswa}

\end{document}